\documentclass[reprint,superscriptaddress,aps,prl]{revtex4-2}

\usepackage[usenames,dvipsnames]{xcolor}
\usepackage[utf8]{inputenc}
\usepackage{bm}
\usepackage{amsmath}
\usepackage{amsfonts,amsthm}
\usepackage{graphicx,psfrag}
\usepackage[psamsfonts]{amssymb}
\usepackage{natbib}
\usepackage{float}
\usepackage{dsfont}
\usepackage{mathtools}
\usepackage{soul,xcolor}
\usepackage[colorlinks=true,linkcolor=Blue,citecolor=Blue,urlcolor=Blue]{hyperref}

\usepackage[caption=false]{subfig}
\usepackage[capitalise]{cleveref}

\setstcolor{magenta}

\DeclareMathOperator*{\argmin}{arg\,min}

\begin{document}

\title{Tempological Control of Network Dynamics}

\author{Yuanzhao Zhang}
\email{yzhang@santafe.edu}
\affiliation{Santa Fe Institute, 1399 Hyde Park Road, Santa Fe, NM 87501, USA}

\author{Sean P. Cornelius}
\email{cornelius@torontomu.ca}
\affiliation{Department of Physics, Toronto Metropolitan University, Toronto, ON, M5B 2K3, Canada}

\begin{abstract} 
    Feedback control is an effective strategy for stabilizing a desired state and has been widely adopted in maintaining the stability of systems such as flying birds and power grids. By default, this framework requires continuous control input to offset deviations from the desired state, which can be invasive and cost a considerable amount of energy. Here, we introduce tempological (temporal + topological) control---a novel, noninvasive strategy that harnesses the inherent flexibility of time-varying networks to control the dynamics of a general nonlinear system.  By strategically switching network topology on-the-fly based on the current states of the nodes, we show how it is possible to drive a system to a desired state or even stabilize otherwise unstable states, all in the absence of external forcing. We demonstrate the utility of our approach using networks of Kuramoto and Stuart-Landau oscillators, achieving synchronization out of sets of unsynchronizable networks. Finally, we develop a statistical theory that explains why tempological control will almost always succeed in the limit of large and diverse temporal networks, with diversity of network configurations overcoming the deficiencies of any snapshot in isolation.
\end{abstract}

\maketitle


Consider one of your favorite networks. Whether of social, biological, or technological nature, there is a good chance that it does not remain static for long.
These dynamic networks with time-varying connections are best described by temporal networks \cite{holme2012temporal,masuda2016guide}, and the consequences of this temporality have been explored from both structural \cite{pan2011path,lentz2013unfolding,pfitzner2013betweenness,peixoto2017modelling,taylor2017eigenvector,paranjape2017motifs,masuda2019detecting} and dynamical \cite{rock2023temporality,belykh2004blinking,stilwell2006sufficient,amritkar2006synchronized,starnini2012random,masuda2013temporal,scholtes2014causality,schroder2015transient,valdano2015analytical,jeter2015synchronization,petit2017theory} perspectives.
For example, previous studies have found that temporal networks hold fundamental advantages over static networks in terms of controllability \cite{li2017fundamental} and synchronizability \cite{zhang2021designing}.  

In many cases, network temporality is not exogenous; instead dynamics \emph{of} the network responds to dynamics \emph{on} the network (often referred to as adaptive or co-evolving networks \cite{gross2008adaptive,sawicki2023perspectives,berner2023adaptive}). 
In the simplest case, the network topology alternates between different static snapshots depending on the system state. For example, in power systems, transmission lines may switch from ``on'' to ``off'' when the current exceeds a certain threshold (line-tripping) \cite{motter2002cascade,yang2017small}. 
Likewise, in social networks, social ties can be severed in response to an epidemic unfolding on the network (social distancing) \cite{gross2006epidemic,block2020social}. 

Dynamics on adaptive networks has been attracting increasing attention in both nonlinear dynamics and network science communities \cite{sorrentino2008adaptive,lu2009adaptive,berner2020birth,berner2021desynchronization}.
However, with a few notable exceptions \cite{lehnert2014controlling,schroder2016interaction}, the possibility of utilizing network temporality as an explicit and adaptive control mechanism has been relatively underexplored.

When controlling networked systems, a common assumption is that nodal dynamics are linear \cite{liu2011controllability,yan2012controlling,yuan2013exact,menichetti2014network,gao2014target,liu2016control,lynn2019physics,baggio2021data}, which allows one to derive elegant controllability conditions using analytical tools from control theory and graph theory.
However, nonlinearity can fundamentally change the controllability of a system \cite{cowan2012nodal,gates2016control}, which motivated the development of various tools for controlling nonlinear networks \cite{kiss2007engineering,liu2011cluster,sun2013controllability,cornelius2013realistic,wells2015control,whalen2015observability,wang2016geometrical,zanudo2017structure,sun2017closed,hart2019topological,morrison2020nonlinear,menara2022functional,sanhedrai2022reviving,d2023controlling}.
In general, analytical results are much harder to establish for nonlinear network control, and the developed approaches are also more varied and diverse.

In this Letter, we introduce a simple and effective control scheme that brings out the potential of temporal networks in steering a coupled system towards desired states, which applies to both linear and nonlinear nodal dynamics.
Unlike previous studies---in which the evolution of the network and the node dynamics are largely decoupled (e.g., as in Ref.~\cite{zhang2021designing})---our scheme makes explicit use of the current states of the nodes, which makes it highly effective even when the system is far from the desired state.
In particular, we show that by strategically switching between available network configurations, one can consistently achieve synchronization from almost any initial conditions, even when no individual network can sustain synchronization on its own.


\emph{Results.} The basic idea behind tempological control is that even when \emph{all} available network configurations fail to produce the desired long-term behavior, some may bring the system closer to it---at least temporarily.

To formalize this strategy, consider a network dynamical system on $n$ nodes, described by:
\begin{equation}
    \dot{x_i} = f_i(x_i) + \sum_{j=1}^n A_{ij}^{\sigma(t)} g(x_i,x_j),     
\end{equation}
where $x_i$ is the state of the $i$th node, $f_i$ its uncoupled dynamics, and $g$ is the interaction function. We assume that at any given time, the network can be in one of $m$ predetermined configurations or \emph{snapshots}, represented by the adjacency matrices $\{\bm{A}^{(1)}, \cdots, \bm{A}^{(m)}\}$. The \emph{switching signal} $\sigma(t) \in \lbrace 1,\ldots,m\rbrace$ represents our control influence on the system. Our objective is to choose $\sigma(t)$---that is, strategically switch between the available network configurations---so as to steer the system towards a desired \emph{target} state $\bm{x}^*=(x_1^*,\cdots,x_n^*)^T$.

\begin{figure*}[t]
\centering
\includegraphics[scale=0.9]{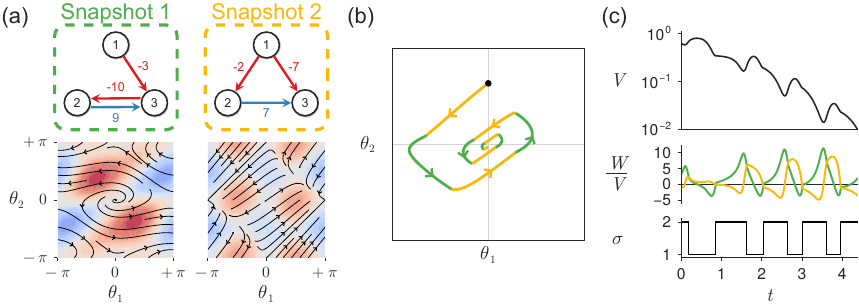}
\vspace{-3mm}
\caption{\textbf{Illustrative example of tempological control.} (a, top) Two snapshots of 3-node networks coupling Kuramoto oscillators. The number beside each directed link $j \rightarrow i$ denotes the corresponding weight $A_{ij}$, and the links are colored blue (red) if $A_{ij}$ is positive (negative). (a, bottom) Vector fields of the corresponding dynamics [\cref{eq:kuramoto}], which we present in 2D by measuring the phases of the first two nodes ($\theta_1$ and $\theta_2$) relative to the third. In both snapshots, the phase-locked state at the origin ($\theta_1 = \theta_2 = \theta_3$) is unstable. However, by strategically switching between these two unstable systems, one can drive the trajectory to the synchronous state from far away, as shown in (b), where we color the controlled trajectory according to which snapshot is active. In (c), we show the corresponding energy ($V$, top), the normalized works of both snapshots ($W/V$, middle), and the switching signal ($\sigma$, bottom) as functions of time.}
\label{fig:schematic}
\end{figure*}

To achieve this, we imagine there is a smooth \emph{energy function} $V(\bm{x})$, with a unique (hence global) minimum $V(\bm{x})=0$ that is attained if and only if $\bm{x} = \bm{x}^*$.
Our choice of the snapshot at time $t$ is based on the rate of change of the energy, $W^{(k)}(\bm{x}) = \dot{V}(\bm{x}) = \nabla V \cdot \dot{\bm{x}}$.
We refer to $W^{(k)}$ as the \emph{work} of snapshot $k$, and simply choose the snapshot with the most negative work for each time $t$:
\begin{equation}
\sigma(t) \;\equiv\; \argmin_k \, W^{(k)}\left(\bm{x}\left(t\right)\right),
\end{equation}
and define the corresponding minimal work as $W_\text{min} = W^{\sigma(t)}$.  In practice, we allow changing to a different configuration after a minimum interval of $\Delta t = 0.05$. The value of $\Delta t$ thus controls the maximum switching rate, and we obtain similar results so long as $\Delta t$ is not too large.
 
As a concrete and canonical example, we consider the following setup based on Kuramoto oscillators:
\begin{equation}
    \dot{\theta}_i = \omega + \sum_{j=1}^n A_{ij}^{\sigma(t)} \sin(\theta_j-\theta_i), \quad i=1,\cdots,n.
    \label{eq:kuramoto}
\end{equation}
We illustrate our control scheme by using it to synchronize otherwise unsynchronizable Kuramoto oscillators (\emph{i.e.}, when all snapshots have substantial repulsive couplings, $A_{ij}^{(k)} < 0$).
Achieving synchronization is important for the function of numerous real-world systems, such as laser arrays \cite{roy1994experimental}, power grids \cite{motter2013spontaneous}, and circadian clocks \cite{zhangenergy,huang2023minimal}.
In this case, it is natural to define the energy function as
\begin{equation}
    V(\bm{\theta}) = \frac{1}{2} - \frac{1}{2n^2}\sum_{i,j} \cos(\theta_j-\theta_i),
    \label{eq:energy}
\end{equation}
whose unique  minimum $V(\bm{\theta})=0$ is achieved only when all oscillators are phase-synchronized ($\theta_1 = \theta_2 = \cdots = \theta_n$) \cite{kuramotoorderparameter}.
This in turn gives the work as
\begin{equation}
     \dot{V}(\bm{\theta}) = \frac{1}{n^2} \sum_{i=1}^n \dot{\theta}_i \left(\sum_{j=1}^n \sin(\theta_i-\theta_j)\right).
    \label{eq:work}
\end{equation}
Combining \cref{eq:kuramoto,eq:work} allows us to calculate $W$ explicitly ($\omega$ can be set to $0$ without loss of generality).
Note that although $V(\bm{\theta})$ only depends on $\bm{\theta}$, $\dot{V}(\bm{\theta})$ also depends on $A_{ij}^{\sigma(t)}$ due to the presence of $\dot{\theta}_i$ in \cref{eq:work}.
We stress that \cref{eq:energy} is not the only valid choice for $V$; any smooth function with a unique minimum at the target can serve.
For example, in phase spaces without periodic boundary conditions, a natural choice is simply
$V(\bm{x}) = \sum_{i=1}^n (x_i-x_i^*)^2$.
In particular, if the target state is full phase synchrony, the energy function can be the synchronization error (see Supplemental Material \cite{SM} for an example).

\Cref{fig:schematic} illustrates the essence of our approach with a small system of $m = 2$ snapshots on $n = 3$ Kuramoto oscillators. Inspection of the vector fields reveals that under the dynamics of \cref{eq:kuramoto}, the synchronous state is unstable in both. 
Accordingly, phase synchronization will never occur in either static snapshot alone. 
Yet,  by switching between these two unstable configurations, one can nonetheless drive the system to synchrony, and keep it there (\cref{fig:schematic}b).
In \cref{fig:schematic}c, we show the show work of each snapshot along the controlled trajectory, and how this gives rise to the switching signal $\sigma(t)$, with switches occurring when one snapshot becomes more favorable than the other.
Interestingly $W_\text{min} > 0$ for some times $t$, and hence the energy $V$ is not monotonically decreasing (\cref{fig:schematic}c, top).
Despite this, the system is successfully driven to synchrony over the long run, even with rather infrequent switching between just two snapshots.

Next, we demonstrate the power of tempological control on an ensemble of synthetic networks.
Each snapshot is generated randomly and independently, where two nodes are connected with probability $q$. 
Among these connections, with probability $p$ the coupling strength is chosen randomly and uniformly in the interval $[-1,0]$, and with probability $1-p$ the coupling strength is chosen randomly and uniformly in the interval $[0,1]$.
Accordingly, $p$ controls how unstable each network configuration is. Our undirected networks thus have an average degree $\langle k \rangle = q\left(n-1\right)$, with the associated weights (including zeros) having mean $\mu = q\left(p - \frac{1}{2}\right)$ and variance $\sigma^2 = q/3 - \mu^2$.

To make the control problem non-trivial, we further require the minimum eigenvalue of the Laplacian matrix $\bm{L}^{(k)} = \bm{D}^{(k)} - \bm{A}^{(k)}$ be sufficiently negative ($<-0.001$) for all snapshots $k = 1,\ldots,m$. 
Here, $\bm{D}^{(k)}$ is a diagonal matrix whose $j$th diagonal entry is given by the $j$th rowsum of the corresponding adjacency matrix $\bm{A}^{(k)}$. 
A negative minimum eigenvalue here translates to a positive largest Lyapunov exponent, guaranteeing the instability of the target state under any static network configuration.

\begin{figure}[tb]
\centering
\subfloat[]{
\includegraphics[width=.95\columnwidth]{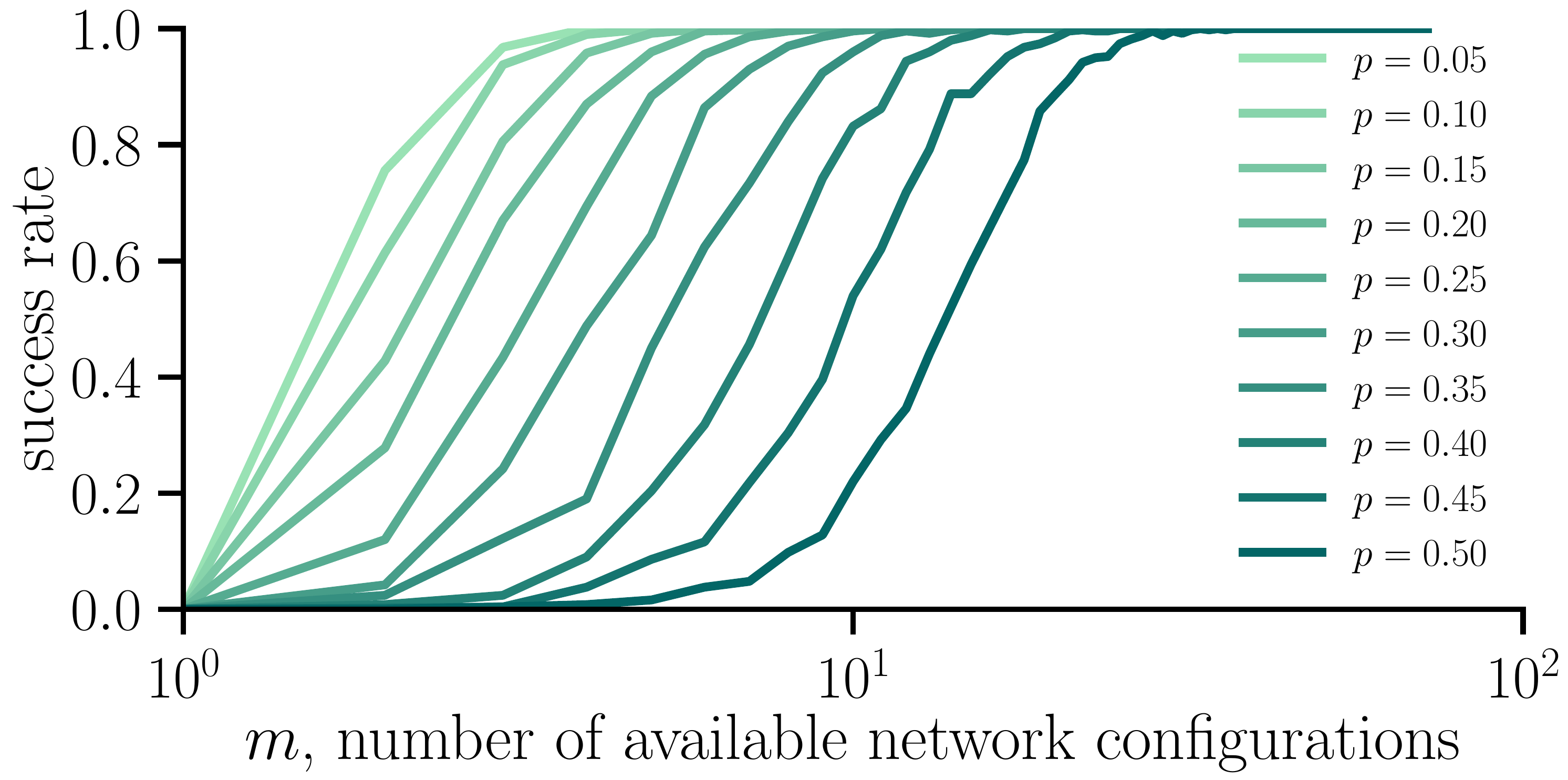}
}
\vspace{-5mm}
\caption{\textbf{Tempological control almost always succeeds with enough network snapshots.}
The success rate $R$ measures the probability of reaching the target synchronization state from random initial conditions by switching between $m$ unstable networks using our tempological control strategy. Each data point is averaged over $500$ independent trials. Here, a trial is considered successful if the energy function $V(\bm{\theta})$ goes below $10^{-6}$ within $300$ time units. The parameter $p$ is the percentage of repulsive links in the networks. Larger $p$ makes the snapshots more unstable, increasing the difficulty of the control task.}
\label{fig:kuramoto_success}
\end{figure}

\Cref{fig:kuramoto_success} shows the control success rate vs.\ the number of snapshots $m$ for various percentages of repulsive links. Here, each snapshot has $n=10$ nodes and we set the connection probability to $q=0.44$ (similar results are obtained for other choices of $n$ and $q$).
With $m=1$, we see that systems never synchronize, as any individual snapshot is (by design) unstable. 
As more network configurations become available, there is more flexibility in choosing a beneficial snapshot. 
Indeed, we see the success rate in reaching the target approaches $1$ for large enough $m$, regardless of the percentages of repulsive links $p$. This approach to $100\%$ success rate is more rapid with smaller $p$, as there is weaker instability to overcome.

We note that our control strategy is not limited to Kuramoto oscillators.
In the Supplemental Material \cite{SM}, we demonstrate the versatility of tempological control by applying it to Stuart-Landau oscillators. 
As in the case of the Kuramoto oscillators presented thus far, we find that our approach is highly effective in synchronizing unstable networks of Stuart-Landau oscillators even with modest numbers of available snapshots. 

\begin{figure}[tb]
\centering
\includegraphics{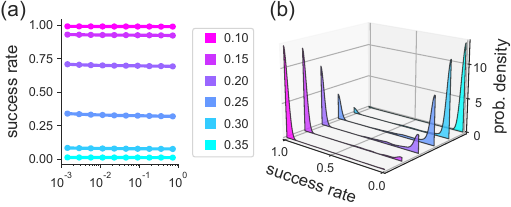}
\vspace{-4mm}
\caption{\textbf{Successful control depends strongly on the available network snapshots, not the initial conditions.}
We attempt to control an ensemble of 1,000 initial conditions towards synchrony, generated to have energies distributed log-uniformly over the range $0.001 \le V_0 \le 0.99$. 
For each value of $p$ (percentage of repulsive links) and each initial condition, we test 1,000 realizations of temporal networks on $n = 20$ nodes, each with $m = 10$ undirected snapshots and a density $q$ corresponding to $\langle k \rangle = 2$. 
(a) The success rate of tempological control is largely independent of $V_0$; states initially far from synchrony ($V_0 \approx 1$) are on average no harder to control than those close to it ($V_0 \approx 0$). (b) In contrast, the success rates of different network realizations are strongly bimodal at $0\%$ and $100\%$. That is, a given set of snapshots will either succeed for almost all initial conditions, or fail for almost all. Changing $p$ merely changes the proportion of temporal networks at either of these extremes.}
\label{fig:success_vs_dist}
\end{figure}

Another advantage of tempological control is that it works for a wide range of different initial conditions, even those far from the target.
This is demonstrated in \cref{fig:success_vs_dist}a, where we plot the success rate versus the energy of the initial condition, $V_0 \equiv V(0)$. For any given percentage of repulsive links $p$, the success rate is essentially constant over a wide range of initial energies $V_0$.
What factors, then, determine the success rate of tempological control?

\Cref{fig:success_vs_dist}b shows that the specific realization of network configurations strongly dictates success versus failure.
Even for the same value of $p$, one realization can be successful for almost all initial conditions, while another realization fails completely for most initial conditions, and the distribution of success rates over different sets fo snapshots is strongly bimodal.

In the limit of large networks with many snapshots, we next argue that \emph{almost all} initial conditions can be brought to the target state under tempological control. 
First, suppose that we have a set of snapshots that covers the phase space with negative work: $W_\text{min}\left(\bm{x}\right) < 0$ whenever $\nabla V \neq \bm{0}$. In this case, the dynamics from any initial condition will asymptotically approach the subspace $\nabla V = \bm{0}$, which contains the target state ($\bm{x}^*$) but possibly other states as well \cite{gradVfootnote}. But by our assumption that the target is the \emph{unique} local minimum of $V$ (which we have the freedom to design), all other stationary states will be saddles or maxima of $V$. Those states are expected to be rendered unstable under our greedy switching dynamics, implying convergence only to $\bm{x}^*$, except on a set of measure zero.

The preceding argument implies that global controllability emerges once we have $W_\text{min} < 0$ over almost the entire phase space. Our task is then to explain the conditions under which this occurs. To make progress toward this goal, 
consider any specific point in phase space $\bm{x}$ for which $\nabla V(\bm{x}) \neq 0$. Note that for any snapshot $k$, the work at $\bm{x}$ is a weighted sum over the (random) $A_{ij}$
\label{eq:worksum}
\begin{equation}
W^{(k)}\left(\bm{x}\right) = \bm{\nabla}V\left(\bm{x}\right) \cdot \bm{f}\left(\bm{x}\right) + \sum_{i, j > i} s_{ij}\left(\bm{x}\right) A_{ij}^{(k)},
\end{equation}
where $s_{ij}\left(\bm{x}\right) \equiv \frac{\partial V}{\partial x_i}g\left(x_i, x_j\right) + \frac{\partial V}{\partial x_j} g\left(x_j, x_i\right)$. We consider undirected networks ($A_{ij} = A_{ij}$) here, hence the sum is only over $j > i$, though the generalization to directed networks is straightforward. 

Now, suppose the $A_{ij}$ are drawn independently from a distribution with mean $\mu$ and finite variance $\sigma^2$. Then by the Lyapunov central limit theorem \cite{billingsley2017probability}, $W\left(\bm{x}\right)$ will be distributed approximately normally as $n \rightarrow \infty$, with mean and variance 
\begin{eqnarray}
\hat{\mu}\left(\bm{x}\right) &=& \bm{\nabla}V\left(\bm{x}\right) \cdot \bm{f}\left(\bm{x}\right) + \mu \sum_{i,j>i} s_{ij}\left(\bm{x}\right), \\
\hat{\sigma}^2\left(\bm{x}\right) &=& \sigma^2 \sum_{i,j>i} s_{ij}^2\left(\bm{x}\right),
\end{eqnarray}
respectively.
Hereafter, we omit the dependence of $\hat{\mu}$, $\hat{\sigma}$ and $W$ on $\bm{x}$ for concision.

We now ask: what will be the \emph{minimal} work over $m$ independent snapshots, $W_\text{min} \equiv \text{min}\left\{W^{(1)},\ldots,W^{(m)}\right\}$?
Under extreme value statistics \cite{fisher1928limiting}, as $m \rightarrow \infty$, $W_\text{min}$ will approximately follow a reverse Gumbel distribution, defined by the cumulative distribution function $C(w) = \mathbb{P}\left(W_\text{min} < w\right)$ given by
\begin{equation}
C\left(w\right) = 
1 - \exp{\left[-\exp{\left(\frac{b_m+\left(w-\hat{\mu}\right)/\hat{\sigma}}{a_m}\right)}\right]}.
\label{eq:gumbel}
\end{equation}
Here, $b_m$ and $a_m > 0$ are constants that depend on the number of snapshots. Through quantile matching \cite{david2004order}, one can determine that $b_m = \sqrt{2}\,\text{erf}^{-1}\left(1-2/m\right)$ and $a_m = \sqrt{2}\,\text{erf}^{-1}\left(2e^{-1/em}-1\right) - b_m$, where $\text{erf}^{-1}$ is the inverse error function.

\Cref{eq:gumbel} makes several key predictions. The statistics of $W_\text{min}$ are, unsurprisingly, influenced by the distribution of link weights, the form of the dynamics, and the location in phase space---all encoded in the parameters $\hat{\mu}$ and $\hat{\sigma}$. More surprising is the presence of another term $\left(b_m/a_m\right)$, which depends only on the \emph{number} of available snapshots, not their specifics. Invoking the asymptotic behavior of $\text{erf}^{-1}$, we predict that as $m \rightarrow \infty$, the distribution of $W_\text{min}$ shifts more negative ($b_m \rightarrow \infty$) and simultaneously becomes narrower ($a_m \rightarrow 0$). Hence, with more and more snapshots, we are increasingly confident that $W_\text{min}$ will be negative, \emph{regardless} of the system specifics.

\begin{figure}[t]
\centering
\includegraphics[width=\columnwidth]{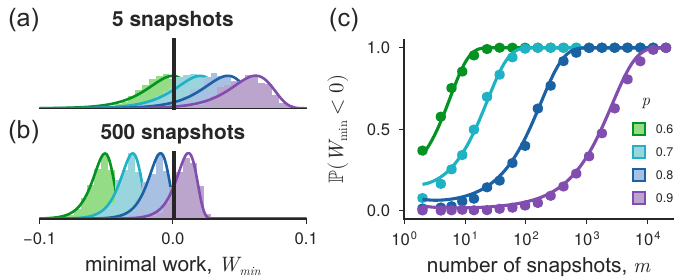}
\vspace{-3mm}
\caption{\textbf{Predicting the distribution and scaling of the minimal work, $W_\text{min}$.} We compare the distribution of $W_\text{min}$ for undirected 100-node temporal networks with: (a) $m = 5$ snaphshots vs. (b) $m = 500$ snapshots, generated with varying $p$ and with $\langle k \rangle = 3$. Solid lines show the density corresponding to our theoretically-predicted Gumbel distribution of \cref{eq:gumbel}, while each set of bars is a histogram over 1,000 random temporal networks with the given parameters. Vertical axes represent density, with the same scale across panels to facilitate comparison. (c) Tail probability of $W_\textrm{min}$ being negative vs.\ the number of snapshots, $m$. Solid lines denote the theoretical predictions of \cref{eq:gumbel} while each marker represents an average over 1,000 random temporal networks.
All simulations consider $W$ evaluated at the initial condition $\bm{\theta}_0$ defined by $\theta_{i + 1} - \theta_i = 0.05$ for $i = 1,\ldots,99$,
with energy $V_0 \approx 0.943$. Other states show similar qualitative behavior, and a similarly excellent fit with theory.}
\label{fig:theory}
\end{figure}

\Cref{fig:theory} confirms our theoretical predictions in simulations on 100-node networks at a representative state $\bm{\theta}_0$. We show empirically-realized distributions of $W_\text{min}$ for $m = 5$ (\cref{fig:theory}a) vs.~$m = 500$ \cref{fig:theory}b snapshots, in both cases an impeccable fit to \cref{eq:gumbel}. With only $m = 5$ snapshots, it is unlikely that the work at $\bm{\theta}_0$ will be negative, unless the percentage of destabilizing links ($p$) is sufficiently low. But with $m = 500$ snapshots, it's a different story. As predicted, the probability mass of $W_\text{min}$ becomes more tightly concentrated at negative values for all $p$.  The odds are thus good to get $W_\text{min} < 0$ even at the largest $p$ considered ($\approx 18.9\%$ at $p = 0.9$), with success virtually guaranteed for lower values of $p$ (e.g., $\approx 96.8\%$ at $p = 0.8$).  This is further confirmed in \cref{fig:theory}c, where we compare fraction of simulations in which $W_\text{min} < 0$ with with the left tail probability predicted by \cref{eq:gumbel}, namely $C(0)$. We see that our theory is an excellent match with numerics, even at moderate values of $m$. Notably, as $m$ becomes large, the probability that $W_\text{min} < 0$ approaches 100\% regardless of $p$. As a consequence, finding a favorable snapshot---and thereby decreasing the energy at $\bm{\theta}_0$---is not a question of if, but when.

\emph{Discussion.} We have shown how one can control the nonlinear dynamics of a network system using temporality (and temporality alone). Our approach is simple, yet 100\% effective in synchronizing Kuramoto and Stuart-Landau oscillators given enough snapshots. We note that tempological control is not limited to synchronization nor to fixed-point targets. Indeed, our approach can be applied to a wide range of target states/nonlinear dynamics; one only needs to construct an appropriate energy function with a unique minimum at the target state.

The energy functions in tempological control differ from Lyapunov functions, both operationally and philosophically. In Lyapunov analysis, the goal is to \emph{certify} the global stability of a state  by identifying an appropriate monotonically decreasing function.
By contrast, in tempological control, the goal is to \emph{engineer} the stability of said state; the energy function is chosen beforehand and informs the control strategy that accomplishes this. Indeed, our $V$ need not even be monotonically decreasing over time; as shown in \cref{fig:schematic}, a system can be steered asymptotically to the target despite occasional increases in energy. This makes it much easier to design an energy function for tempological control compared to a Lyapunov function.

We also contrast our approach with the traditional switching control literature, the lion's share of which focuses on linear snapshot dynamics for the sake of analytical tractability \cite{lin2009stability}. For the comparatively few existing works on nonlinear switching systems, the control signal needs to be re-calculated in advance for each new initial condition \cite{axelsson2008optimal}. Our strategy applies to nonlinear dynamics and can be executed on-the-fly without prior planning---both valuable traits in practical applications.

More generally, our approach falls within the broad category of closed-loop (feedback) control, which offers several advantages over open-loop schemes.
Open-loop schemes are generally easier to implement, require fewer adjustments when applied to different systems, and are unencumbered by complications such as measurement errors and delays.
However, by taking the current states of the system into account, closed-loop schemes can achieve tasks beyond the reach of open-loop schemes.
We confirm this in the Supplemental Material \cite{SM}, where we compare tempological control with a simple open-loop strategy of randomly switching among the available network configurations. We see that optimizing when and which snapshot to switch to consistently boosts the success rate, especially for networks with a high percentage of repulsive links.
We hope this work will inspire additional novel control strategies in the future for nonlinear, high-dimensional systems with complex network structures.

\begin{acknowledgments}
\emph{Acknowledgments.} YZ acknowledges support from the Omidyar Fellowship and the National Science Foundation (NSF DMS 2436231). SPC acknowledges support from the Natural Science and Engineering Research Council of Canada (NSERC) and the Digital Research Alliance of Canada.
\end{acknowledgments}

\bibliography{ref}

\begin{thebibliography}{71}%
\makeatletter
\providecommand \@ifxundefined [1]{%
 \@ifx{#1\undefined}
}%
\providecommand \@ifnum [1]{%
 \ifnum #1\expandafter \@firstoftwo
 \else \expandafter \@secondoftwo
 \fi
}%
\providecommand \@ifx [1]{%
 \ifx #1\expandafter \@firstoftwo
 \else \expandafter \@secondoftwo
 \fi
}%
\providecommand \natexlab [1]{#1}%
\providecommand \enquote  [1]{``#1''}%
\providecommand \bibnamefont  [1]{#1}%
\providecommand \bibfnamefont [1]{#1}%
\providecommand \citenamefont [1]{#1}%
\providecommand \href@noop [0]{\@secondoftwo}%
\providecommand \href [0]{\begingroup \@sanitize@url \@href}%
\providecommand \@href[1]{\@@startlink{#1}\@@href}%
\providecommand \@@href[1]{\endgroup#1\@@endlink}%
\providecommand \@sanitize@url [0]{\catcode `\\12\catcode `\$12\catcode
  `\&12\catcode `\#12\catcode `\^12\catcode `\_12\catcode `\%12\relax}%
\providecommand \@@startlink[1]{}%
\providecommand \@@endlink[0]{}%
\providecommand \url  [0]{\begingroup\@sanitize@url \@url }%
\providecommand \@url [1]{\endgroup\@href {#1}{\urlprefix }}%
\providecommand \urlprefix  [0]{URL }%
\providecommand \Eprint [0]{\href }%
\providecommand \doibase [0]{https://doi.org/}%
\providecommand \selectlanguage [0]{\@gobble}%
\providecommand \bibinfo  [0]{\@secondoftwo}%
\providecommand \bibfield  [0]{\@secondoftwo}%
\providecommand \translation [1]{[#1]}%
\providecommand \BibitemOpen [0]{}%
\providecommand \bibitemStop [0]{}%
\providecommand \bibitemNoStop [0]{.\EOS\space}%
\providecommand \EOS [0]{\spacefactor3000\relax}%
\providecommand \BibitemShut  [1]{\csname bibitem#1\endcsname}%
\let\auto@bib@innerbib\@empty
\bibitem [{\citenamefont {Holme}\ and\ \citenamefont
  {Saram{\"a}ki}(2012)}]{holme2012temporal}%
  \BibitemOpen
  \bibfield  {author} {\bibinfo {author} {\bibfnamefont {P.}~\bibnamefont
  {Holme}}\ and\ \bibinfo {author} {\bibfnamefont {J.}~\bibnamefont
  {Saram{\"a}ki}},\ }\bibfield  {title} {\bibinfo {title} {Temporal networks},\
  }\href@noop {} {\bibfield  {journal} {\bibinfo  {journal} {Phys. Rep.}\
  }\textbf {\bibinfo {volume} {519}},\ \bibinfo {pages} {97} (\bibinfo {year}
  {2012})}\BibitemShut {NoStop}%
\bibitem [{\citenamefont {Masuda}\ and\ \citenamefont
  {Lambiotte}(2016)}]{masuda2016guide}%
  \BibitemOpen
  \bibfield  {author} {\bibinfo {author} {\bibfnamefont {N.}~\bibnamefont
  {Masuda}}\ and\ \bibinfo {author} {\bibfnamefont {R.}~\bibnamefont
  {Lambiotte}},\ }\href@noop {} {\emph {\bibinfo {title} {A guide to temporal
  networks}}}\ (\bibinfo  {publisher} {World Scientific},\ \bibinfo {year}
  {2016})\BibitemShut {NoStop}%
\bibitem [{\citenamefont {Pan}\ and\ \citenamefont
  {Saram{\"a}ki}(2011)}]{pan2011path}%
  \BibitemOpen
  \bibfield  {author} {\bibinfo {author} {\bibfnamefont {R.~K.}\ \bibnamefont
  {Pan}}\ and\ \bibinfo {author} {\bibfnamefont {J.}~\bibnamefont
  {Saram{\"a}ki}},\ }\bibfield  {title} {\bibinfo {title} {Path lengths,
  correlations, and centrality in temporal networks},\ }\href@noop {}
  {\bibfield  {journal} {\bibinfo  {journal} {Phys. Rev. E}\ }\textbf {\bibinfo
  {volume} {84}},\ \bibinfo {pages} {016105} (\bibinfo {year}
  {2011})}\BibitemShut {NoStop}%
\bibitem [{\citenamefont {Lentz}\ \emph {et~al.}(2013)\citenamefont {Lentz},
  \citenamefont {Selhorst},\ and\ \citenamefont
  {Sokolov}}]{lentz2013unfolding}%
  \BibitemOpen
  \bibfield  {author} {\bibinfo {author} {\bibfnamefont {H.~H.}\ \bibnamefont
  {Lentz}}, \bibinfo {author} {\bibfnamefont {T.}~\bibnamefont {Selhorst}},\
  and\ \bibinfo {author} {\bibfnamefont {I.~M.}\ \bibnamefont {Sokolov}},\
  }\bibfield  {title} {\bibinfo {title} {Unfolding accessibility provides a
  macroscopic approach to temporal networks},\ }\href@noop {} {\bibfield
  {journal} {\bibinfo  {journal} {Phys. Rev. Lett.}\ }\textbf {\bibinfo
  {volume} {110}},\ \bibinfo {pages} {118701} (\bibinfo {year}
  {2013})}\BibitemShut {NoStop}%
\bibitem [{\citenamefont {Pfitzner}\ \emph {et~al.}(2013)\citenamefont
  {Pfitzner}, \citenamefont {Scholtes}, \citenamefont {Garas}, \citenamefont
  {Tessone},\ and\ \citenamefont {Schweitzer}}]{pfitzner2013betweenness}%
  \BibitemOpen
  \bibfield  {author} {\bibinfo {author} {\bibfnamefont {R.}~\bibnamefont
  {Pfitzner}}, \bibinfo {author} {\bibfnamefont {I.}~\bibnamefont {Scholtes}},
  \bibinfo {author} {\bibfnamefont {A.}~\bibnamefont {Garas}}, \bibinfo
  {author} {\bibfnamefont {C.~J.}\ \bibnamefont {Tessone}},\ and\ \bibinfo
  {author} {\bibfnamefont {F.}~\bibnamefont {Schweitzer}},\ }\bibfield  {title}
  {\bibinfo {title} {Betweenness preference: Quantifying correlations in the
  topological dynamics of temporal networks},\ }\href@noop {} {\bibfield
  {journal} {\bibinfo  {journal} {Phys. Rev. Lett.}\ }\textbf {\bibinfo
  {volume} {110}},\ \bibinfo {pages} {198701} (\bibinfo {year}
  {2013})}\BibitemShut {NoStop}%
\bibitem [{\citenamefont {Peixoto}\ and\ \citenamefont
  {Rosvall}(2017)}]{peixoto2017modelling}%
  \BibitemOpen
  \bibfield  {author} {\bibinfo {author} {\bibfnamefont {T.~P.}\ \bibnamefont
  {Peixoto}}\ and\ \bibinfo {author} {\bibfnamefont {M.}~\bibnamefont
  {Rosvall}},\ }\bibfield  {title} {\bibinfo {title} {Modelling sequences and
  temporal networks with dynamic community structures},\ }\href@noop {}
  {\bibfield  {journal} {\bibinfo  {journal} {Nat. Commun.}\ }\textbf {\bibinfo
  {volume} {8}},\ \bibinfo {pages} {582} (\bibinfo {year} {2017})}\BibitemShut
  {NoStop}%
\bibitem [{\citenamefont {Taylor}\ \emph {et~al.}(2017)\citenamefont {Taylor},
  \citenamefont {Myers}, \citenamefont {Clauset}, \citenamefont {Porter},\ and\
  \citenamefont {Mucha}}]{taylor2017eigenvector}%
  \BibitemOpen
  \bibfield  {author} {\bibinfo {author} {\bibfnamefont {D.}~\bibnamefont
  {Taylor}}, \bibinfo {author} {\bibfnamefont {S.~A.}\ \bibnamefont {Myers}},
  \bibinfo {author} {\bibfnamefont {A.}~\bibnamefont {Clauset}}, \bibinfo
  {author} {\bibfnamefont {M.~A.}\ \bibnamefont {Porter}},\ and\ \bibinfo
  {author} {\bibfnamefont {P.~J.}\ \bibnamefont {Mucha}},\ }\bibfield  {title}
  {\bibinfo {title} {Eigenvector-based centrality measures for temporal
  networks},\ }\href@noop {} {\bibfield  {journal} {\bibinfo  {journal}
  {Multiscale Model. Simul.}\ }\textbf {\bibinfo {volume} {15}},\ \bibinfo
  {pages} {537} (\bibinfo {year} {2017})}\BibitemShut {NoStop}%
\bibitem [{\citenamefont {Paranjape}\ \emph {et~al.}(2017)\citenamefont
  {Paranjape}, \citenamefont {Benson},\ and\ \citenamefont
  {Leskovec}}]{paranjape2017motifs}%
  \BibitemOpen
  \bibfield  {author} {\bibinfo {author} {\bibfnamefont {A.}~\bibnamefont
  {Paranjape}}, \bibinfo {author} {\bibfnamefont {A.~R.}\ \bibnamefont
  {Benson}},\ and\ \bibinfo {author} {\bibfnamefont {J.}~\bibnamefont
  {Leskovec}},\ }\bibfield  {title} {\bibinfo {title} {Motifs in temporal
  networks},\ }in\ \href@noop {} {\emph {\bibinfo {booktitle} {Proceedings of
  the Tenth ACM International Conference on Web Search and Data Mining}}}\
  (\bibinfo {year} {2017})\ pp.\ \bibinfo {pages} {601--610}\BibitemShut
  {NoStop}%
\bibitem [{\citenamefont {Masuda}\ and\ \citenamefont
  {Holme}(2019)}]{masuda2019detecting}%
  \BibitemOpen
  \bibfield  {author} {\bibinfo {author} {\bibfnamefont {N.}~\bibnamefont
  {Masuda}}\ and\ \bibinfo {author} {\bibfnamefont {P.}~\bibnamefont {Holme}},\
  }\bibfield  {title} {\bibinfo {title} {Detecting sequences of system states
  in temporal networks},\ }\href@noop {} {\bibfield  {journal} {\bibinfo
  {journal} {Sci. Rep.}\ }\textbf {\bibinfo {volume} {9}},\ \bibinfo {pages}
  {795} (\bibinfo {year} {2019})}\BibitemShut {NoStop}%
\bibitem [{\citenamefont {Rock}\ \emph {et~al.}(2023)\citenamefont {Rock},
  \citenamefont {Dirie},\ and\ \citenamefont
  {Cornelius}}]{rock2023temporality}%
  \BibitemOpen
  \bibfield  {author} {\bibinfo {author} {\bibfnamefont {K.~M.}\ \bibnamefont
  {Rock}}, \bibinfo {author} {\bibfnamefont {H.}~\bibnamefont {Dirie}},\ and\
  \bibinfo {author} {\bibfnamefont {S.~P.}\ \bibnamefont {Cornelius}},\
  }\bibfield  {title} {\bibinfo {title} {Temporality-induced chaos in the
  kuramoto model},\ }\href@noop {} {\bibfield  {journal} {\bibinfo  {journal}
  {Northeast Journal of Complex Systems (NEJCS)}\ }\textbf {\bibinfo {volume}
  {5}},\ \bibinfo {pages} {3} (\bibinfo {year} {2023})}\BibitemShut {NoStop}%
\bibitem [{\citenamefont {Belykh}\ \emph {et~al.}(2004)\citenamefont {Belykh},
  \citenamefont {Belykh},\ and\ \citenamefont {Hasler}}]{belykh2004blinking}%
  \BibitemOpen
  \bibfield  {author} {\bibinfo {author} {\bibfnamefont {I.~V.}\ \bibnamefont
  {Belykh}}, \bibinfo {author} {\bibfnamefont {V.~N.}\ \bibnamefont {Belykh}},\
  and\ \bibinfo {author} {\bibfnamefont {M.}~\bibnamefont {Hasler}},\
  }\bibfield  {title} {\bibinfo {title} {Blinking model and synchronization in
  small-world networks with a time-varying coupling},\ }\href@noop {}
  {\bibfield  {journal} {\bibinfo  {journal} {Physica D}\ }\textbf {\bibinfo
  {volume} {195}},\ \bibinfo {pages} {188} (\bibinfo {year}
  {2004})}\BibitemShut {NoStop}%
\bibitem [{\citenamefont {Stilwell}\ \emph {et~al.}(2006)\citenamefont
  {Stilwell}, \citenamefont {Bollt},\ and\ \citenamefont
  {Roberson}}]{stilwell2006sufficient}%
  \BibitemOpen
  \bibfield  {author} {\bibinfo {author} {\bibfnamefont {D.~J.}\ \bibnamefont
  {Stilwell}}, \bibinfo {author} {\bibfnamefont {E.~M.}\ \bibnamefont
  {Bollt}},\ and\ \bibinfo {author} {\bibfnamefont {D.~G.}\ \bibnamefont
  {Roberson}},\ }\bibfield  {title} {\bibinfo {title} {Sufficient conditions
  for fast switching synchronization in time-varying network topologies},\
  }\href@noop {} {\bibfield  {journal} {\bibinfo  {journal} {SIAM J. Appl. Dyn.
  Syst.}\ }\textbf {\bibinfo {volume} {5}},\ \bibinfo {pages} {140} (\bibinfo
  {year} {2006})}\BibitemShut {NoStop}%
\bibitem [{\citenamefont {Amritkar}\ and\ \citenamefont
  {Hu}(2006)}]{amritkar2006synchronized}%
  \BibitemOpen
  \bibfield  {author} {\bibinfo {author} {\bibfnamefont {R.}~\bibnamefont
  {Amritkar}}\ and\ \bibinfo {author} {\bibfnamefont {C.-K.}\ \bibnamefont
  {Hu}},\ }\bibfield  {title} {\bibinfo {title} {Synchronized state of coupled
  dynamics on time-varying networks},\ }\href@noop {} {\bibfield  {journal}
  {\bibinfo  {journal} {Chaos}\ }\textbf {\bibinfo {volume} {16}},\ \bibinfo
  {pages} {015117} (\bibinfo {year} {2006})}\BibitemShut {NoStop}%
\bibitem [{\citenamefont {Starnini}\ \emph {et~al.}(2012)\citenamefont
  {Starnini}, \citenamefont {Baronchelli}, \citenamefont {Barrat},\ and\
  \citenamefont {Pastor-Satorras}}]{starnini2012random}%
  \BibitemOpen
  \bibfield  {author} {\bibinfo {author} {\bibfnamefont {M.}~\bibnamefont
  {Starnini}}, \bibinfo {author} {\bibfnamefont {A.}~\bibnamefont
  {Baronchelli}}, \bibinfo {author} {\bibfnamefont {A.}~\bibnamefont
  {Barrat}},\ and\ \bibinfo {author} {\bibfnamefont {R.}~\bibnamefont
  {Pastor-Satorras}},\ }\bibfield  {title} {\bibinfo {title} {Random walks on
  temporal networks},\ }\href@noop {} {\bibfield  {journal} {\bibinfo
  {journal} {Phys. Rev. E}\ }\textbf {\bibinfo {volume} {85}},\ \bibinfo
  {pages} {056115} (\bibinfo {year} {2012})}\BibitemShut {NoStop}%
\bibitem [{\citenamefont {Masuda}\ \emph {et~al.}(2013)\citenamefont {Masuda},
  \citenamefont {Klemm},\ and\ \citenamefont
  {Egu{\'\i}luz}}]{masuda2013temporal}%
  \BibitemOpen
  \bibfield  {author} {\bibinfo {author} {\bibfnamefont {N.}~\bibnamefont
  {Masuda}}, \bibinfo {author} {\bibfnamefont {K.}~\bibnamefont {Klemm}},\ and\
  \bibinfo {author} {\bibfnamefont {V.~M.}\ \bibnamefont {Egu{\'\i}luz}},\
  }\bibfield  {title} {\bibinfo {title} {Temporal networks: Slowing down
  diffusion by long lasting interactions},\ }\href@noop {} {\bibfield
  {journal} {\bibinfo  {journal} {Phys. Rev. Lett.}\ }\textbf {\bibinfo
  {volume} {111}},\ \bibinfo {pages} {188701} (\bibinfo {year}
  {2013})}\BibitemShut {NoStop}%
\bibitem [{\citenamefont {Scholtes}\ \emph {et~al.}(2014)\citenamefont
  {Scholtes}, \citenamefont {Wider}, \citenamefont {Pfitzner}, \citenamefont
  {Garas}, \citenamefont {Tessone},\ and\ \citenamefont
  {Schweitzer}}]{scholtes2014causality}%
  \BibitemOpen
  \bibfield  {author} {\bibinfo {author} {\bibfnamefont {I.}~\bibnamefont
  {Scholtes}}, \bibinfo {author} {\bibfnamefont {N.}~\bibnamefont {Wider}},
  \bibinfo {author} {\bibfnamefont {R.}~\bibnamefont {Pfitzner}}, \bibinfo
  {author} {\bibfnamefont {A.}~\bibnamefont {Garas}}, \bibinfo {author}
  {\bibfnamefont {C.~J.}\ \bibnamefont {Tessone}},\ and\ \bibinfo {author}
  {\bibfnamefont {F.}~\bibnamefont {Schweitzer}},\ }\bibfield  {title}
  {\bibinfo {title} {Causality-driven slow-down and speed-up of diffusion in
  non-markovian temporal networks},\ }\href@noop {} {\bibfield  {journal}
  {\bibinfo  {journal} {Nat. Commun.}\ }\textbf {\bibinfo {volume} {5}},\
  \bibinfo {pages} {5024} (\bibinfo {year} {2014})}\BibitemShut {NoStop}%
\bibitem [{\citenamefont {Schr{\"o}der}\ \emph {et~al.}(2015)\citenamefont
  {Schr{\"o}der}, \citenamefont {Mannattil}, \citenamefont {Dutta},
  \citenamefont {Chakraborty},\ and\ \citenamefont
  {Timme}}]{schroder2015transient}%
  \BibitemOpen
  \bibfield  {author} {\bibinfo {author} {\bibfnamefont {M.}~\bibnamefont
  {Schr{\"o}der}}, \bibinfo {author} {\bibfnamefont {M.}~\bibnamefont
  {Mannattil}}, \bibinfo {author} {\bibfnamefont {D.}~\bibnamefont {Dutta}},
  \bibinfo {author} {\bibfnamefont {S.}~\bibnamefont {Chakraborty}},\ and\
  \bibinfo {author} {\bibfnamefont {M.}~\bibnamefont {Timme}},\ }\bibfield
  {title} {\bibinfo {title} {Transient uncoupling induces synchronization},\
  }\href@noop {} {\bibfield  {journal} {\bibinfo  {journal} {Phys. Rev. Lett.}\
  }\textbf {\bibinfo {volume} {115}},\ \bibinfo {pages} {054101} (\bibinfo
  {year} {2015})}\BibitemShut {NoStop}%
\bibitem [{\citenamefont {Valdano}\ \emph {et~al.}(2015)\citenamefont
  {Valdano}, \citenamefont {Ferreri}, \citenamefont {Poletto},\ and\
  \citenamefont {Colizza}}]{valdano2015analytical}%
  \BibitemOpen
  \bibfield  {author} {\bibinfo {author} {\bibfnamefont {E.}~\bibnamefont
  {Valdano}}, \bibinfo {author} {\bibfnamefont {L.}~\bibnamefont {Ferreri}},
  \bibinfo {author} {\bibfnamefont {C.}~\bibnamefont {Poletto}},\ and\ \bibinfo
  {author} {\bibfnamefont {V.}~\bibnamefont {Colizza}},\ }\bibfield  {title}
  {\bibinfo {title} {Analytical computation of the epidemic threshold on
  temporal networks},\ }\href@noop {} {\bibfield  {journal} {\bibinfo
  {journal} {Phys. Rev. X}\ }\textbf {\bibinfo {volume} {5}},\ \bibinfo {pages}
  {021005} (\bibinfo {year} {2015})}\BibitemShut {NoStop}%
\bibitem [{\citenamefont {Jeter}\ and\ \citenamefont
  {Belykh}(2015)}]{jeter2015synchronization}%
  \BibitemOpen
  \bibfield  {author} {\bibinfo {author} {\bibfnamefont {R.}~\bibnamefont
  {Jeter}}\ and\ \bibinfo {author} {\bibfnamefont {I.}~\bibnamefont {Belykh}},\
  }\bibfield  {title} {\bibinfo {title} {Synchronization in on-off stochastic
  networks: Windows of opportunity},\ }\href@noop {} {\bibfield  {journal}
  {\bibinfo  {journal} {IEEE Trans. Circuits Syst. I, Reg. Papers}\ }\textbf
  {\bibinfo {volume} {62}},\ \bibinfo {pages} {1260} (\bibinfo {year}
  {2015})}\BibitemShut {NoStop}%
\bibitem [{\citenamefont {Petit}\ \emph {et~al.}(2017)\citenamefont {Petit},
  \citenamefont {Lauwens}, \citenamefont {Fanelli},\ and\ \citenamefont
  {Carletti}}]{petit2017theory}%
  \BibitemOpen
  \bibfield  {author} {\bibinfo {author} {\bibfnamefont {J.}~\bibnamefont
  {Petit}}, \bibinfo {author} {\bibfnamefont {B.}~\bibnamefont {Lauwens}},
  \bibinfo {author} {\bibfnamefont {D.}~\bibnamefont {Fanelli}},\ and\ \bibinfo
  {author} {\bibfnamefont {T.}~\bibnamefont {Carletti}},\ }\bibfield  {title}
  {\bibinfo {title} {Theory of {T}uring patterns on time varying networks},\
  }\href@noop {} {\bibfield  {journal} {\bibinfo  {journal} {Phys. Rev. Lett.}\
  }\textbf {\bibinfo {volume} {119}},\ \bibinfo {pages} {148301} (\bibinfo
  {year} {2017})}\BibitemShut {NoStop}%
\bibitem [{\citenamefont {Li}\ \emph {et~al.}(2017)\citenamefont {Li},
  \citenamefont {Cornelius}, \citenamefont {Liu}, \citenamefont {Wang},\ and\
  \citenamefont {Barab{\'a}si}}]{li2017fundamental}%
  \BibitemOpen
  \bibfield  {author} {\bibinfo {author} {\bibfnamefont {A.}~\bibnamefont
  {Li}}, \bibinfo {author} {\bibfnamefont {S.~P.}\ \bibnamefont {Cornelius}},
  \bibinfo {author} {\bibfnamefont {Y.-Y.}\ \bibnamefont {Liu}}, \bibinfo
  {author} {\bibfnamefont {L.}~\bibnamefont {Wang}},\ and\ \bibinfo {author}
  {\bibfnamefont {A.-L.}\ \bibnamefont {Barab{\'a}si}},\ }\bibfield  {title}
  {\bibinfo {title} {The fundamental advantages of temporal networks},\
  }\href@noop {} {\bibfield  {journal} {\bibinfo  {journal} {Science}\ }\textbf
  {\bibinfo {volume} {358}},\ \bibinfo {pages} {1042} (\bibinfo {year}
  {2017})}\BibitemShut {NoStop}%
\bibitem [{\citenamefont {Zhang}\ and\ \citenamefont
  {Strogatz}(2021)}]{zhang2021designing}%
  \BibitemOpen
  \bibfield  {author} {\bibinfo {author} {\bibfnamefont {Y.}~\bibnamefont
  {Zhang}}\ and\ \bibinfo {author} {\bibfnamefont {S.~H.}\ \bibnamefont
  {Strogatz}},\ }\bibfield  {title} {\bibinfo {title} {Designing temporal
  networks that synchronize under resource constraints},\ }\href@noop {}
  {\bibfield  {journal} {\bibinfo  {journal} {Nat. Commun.}\ }\textbf {\bibinfo
  {volume} {12}},\ \bibinfo {pages} {3273} (\bibinfo {year}
  {2021})}\BibitemShut {NoStop}%
\bibitem [{\citenamefont {Gross}\ and\ \citenamefont
  {Blasius}(2008)}]{gross2008adaptive}%
  \BibitemOpen
  \bibfield  {author} {\bibinfo {author} {\bibfnamefont {T.}~\bibnamefont
  {Gross}}\ and\ \bibinfo {author} {\bibfnamefont {B.}~\bibnamefont
  {Blasius}},\ }\bibfield  {title} {\bibinfo {title} {Adaptive coevolutionary
  networks: A review},\ }\href@noop {} {\bibfield  {journal} {\bibinfo
  {journal} {J. R. Soc. Interface}\ }\textbf {\bibinfo {volume} {5}},\ \bibinfo
  {pages} {259} (\bibinfo {year} {2008})}\BibitemShut {NoStop}%
\bibitem [{\citenamefont {Sawicki}\ \emph {et~al.}(2023)\citenamefont
  {Sawicki}, \citenamefont {Berner}, \citenamefont {Loos}, \citenamefont
  {Anvari}, \citenamefont {Bader}, \citenamefont {Barfuss}, \citenamefont
  {Botta}, \citenamefont {Brede}, \citenamefont {Franovi{\'c}}, \citenamefont
  {Gauthier} \emph {et~al.}}]{sawicki2023perspectives}%
  \BibitemOpen
  \bibfield  {author} {\bibinfo {author} {\bibfnamefont {J.}~\bibnamefont
  {Sawicki}}, \bibinfo {author} {\bibfnamefont {R.}~\bibnamefont {Berner}},
  \bibinfo {author} {\bibfnamefont {S.~A.}\ \bibnamefont {Loos}}, \bibinfo
  {author} {\bibfnamefont {M.}~\bibnamefont {Anvari}}, \bibinfo {author}
  {\bibfnamefont {R.}~\bibnamefont {Bader}}, \bibinfo {author} {\bibfnamefont
  {W.}~\bibnamefont {Barfuss}}, \bibinfo {author} {\bibfnamefont
  {N.}~\bibnamefont {Botta}}, \bibinfo {author} {\bibfnamefont
  {N.}~\bibnamefont {Brede}}, \bibinfo {author} {\bibfnamefont
  {I.}~\bibnamefont {Franovi{\'c}}}, \bibinfo {author} {\bibfnamefont {D.~J.}\
  \bibnamefont {Gauthier}}, \emph {et~al.},\ }\bibfield  {title} {\bibinfo
  {title} {Perspectives on adaptive dynamical systems},\ }\href@noop {}
  {\bibfield  {journal} {\bibinfo  {journal} {Chaos}\ }\textbf {\bibinfo
  {volume} {33}} (\bibinfo {year} {2023})}\BibitemShut {NoStop}%
\bibitem [{\citenamefont {Berner}\ \emph {et~al.}(2023)\citenamefont {Berner},
  \citenamefont {Gross}, \citenamefont {Kuehn}, \citenamefont {Kurths},\ and\
  \citenamefont {Yanchuk}}]{berner2023adaptive}%
  \BibitemOpen
  \bibfield  {author} {\bibinfo {author} {\bibfnamefont {R.}~\bibnamefont
  {Berner}}, \bibinfo {author} {\bibfnamefont {T.}~\bibnamefont {Gross}},
  \bibinfo {author} {\bibfnamefont {C.}~\bibnamefont {Kuehn}}, \bibinfo
  {author} {\bibfnamefont {J.}~\bibnamefont {Kurths}},\ and\ \bibinfo {author}
  {\bibfnamefont {S.}~\bibnamefont {Yanchuk}},\ }\bibfield  {title} {\bibinfo
  {title} {Adaptive dynamical networks},\ }\href@noop {} {\bibfield  {journal}
  {\bibinfo  {journal} {Phys. Rep.}\ }\textbf {\bibinfo {volume} {1031}},\
  \bibinfo {pages} {1} (\bibinfo {year} {2023})}\BibitemShut {NoStop}%
\bibitem [{\citenamefont {Motter}\ and\ \citenamefont
  {Lai}(2002)}]{motter2002cascade}%
  \BibitemOpen
  \bibfield  {author} {\bibinfo {author} {\bibfnamefont {A.~E.}\ \bibnamefont
  {Motter}}\ and\ \bibinfo {author} {\bibfnamefont {Y.-C.}\ \bibnamefont
  {Lai}},\ }\bibfield  {title} {\bibinfo {title} {Cascade-based attacks on
  complex networks},\ }\href@noop {} {\bibfield  {journal} {\bibinfo  {journal}
  {Phys. Rev. E}\ }\textbf {\bibinfo {volume} {66}},\ \bibinfo {pages} {065102}
  (\bibinfo {year} {2002})}\BibitemShut {NoStop}%
\bibitem [{\citenamefont {Yang}\ \emph {et~al.}(2017)\citenamefont {Yang},
  \citenamefont {Nishikawa},\ and\ \citenamefont {Motter}}]{yang2017small}%
  \BibitemOpen
  \bibfield  {author} {\bibinfo {author} {\bibfnamefont {Y.}~\bibnamefont
  {Yang}}, \bibinfo {author} {\bibfnamefont {T.}~\bibnamefont {Nishikawa}},\
  and\ \bibinfo {author} {\bibfnamefont {A.~E.}\ \bibnamefont {Motter}},\
  }\bibfield  {title} {\bibinfo {title} {Small vulnerable sets determine large
  network cascades in power grids},\ }\href@noop {} {\bibfield  {journal}
  {\bibinfo  {journal} {Science}\ }\textbf {\bibinfo {volume} {358}},\ \bibinfo
  {pages} {eaan3184} (\bibinfo {year} {2017})}\BibitemShut {NoStop}%
\bibitem [{\citenamefont {Gross}\ \emph {et~al.}(2006)\citenamefont {Gross},
  \citenamefont {D'Lima},\ and\ \citenamefont {Blasius}}]{gross2006epidemic}%
  \BibitemOpen
  \bibfield  {author} {\bibinfo {author} {\bibfnamefont {T.}~\bibnamefont
  {Gross}}, \bibinfo {author} {\bibfnamefont {C.~J.~D.}\ \bibnamefont
  {D'Lima}},\ and\ \bibinfo {author} {\bibfnamefont {B.}~\bibnamefont
  {Blasius}},\ }\bibfield  {title} {\bibinfo {title} {Epidemic dynamics on an
  adaptive network},\ }\href@noop {} {\bibfield  {journal} {\bibinfo  {journal}
  {Phys. Rev. Lett.}\ }\textbf {\bibinfo {volume} {96}},\ \bibinfo {pages}
  {208701} (\bibinfo {year} {2006})}\BibitemShut {NoStop}%
\bibitem [{\citenamefont {Block}\ \emph {et~al.}(2020)\citenamefont {Block},
  \citenamefont {Hoffman}, \citenamefont {Raabe}, \citenamefont {Dowd},
  \citenamefont {Rahal}, \citenamefont {Kashyap},\ and\ \citenamefont
  {Mills}}]{block2020social}%
  \BibitemOpen
  \bibfield  {author} {\bibinfo {author} {\bibfnamefont {P.}~\bibnamefont
  {Block}}, \bibinfo {author} {\bibfnamefont {M.}~\bibnamefont {Hoffman}},
  \bibinfo {author} {\bibfnamefont {I.~J.}\ \bibnamefont {Raabe}}, \bibinfo
  {author} {\bibfnamefont {J.~B.}\ \bibnamefont {Dowd}}, \bibinfo {author}
  {\bibfnamefont {C.}~\bibnamefont {Rahal}}, \bibinfo {author} {\bibfnamefont
  {R.}~\bibnamefont {Kashyap}},\ and\ \bibinfo {author} {\bibfnamefont {M.~C.}\
  \bibnamefont {Mills}},\ }\bibfield  {title} {\bibinfo {title} {Social
  network-based distancing strategies to flatten the covid-19 curve in a
  post-lockdown world},\ }\href@noop {} {\bibfield  {journal} {\bibinfo
  {journal} {Nat. Hum. Behav.}\ }\textbf {\bibinfo {volume} {4}},\ \bibinfo
  {pages} {588} (\bibinfo {year} {2020})}\BibitemShut {NoStop}%
\bibitem [{\citenamefont {Sorrentino}\ and\ \citenamefont
  {Ott}(2008)}]{sorrentino2008adaptive}%
  \BibitemOpen
  \bibfield  {author} {\bibinfo {author} {\bibfnamefont {F.}~\bibnamefont
  {Sorrentino}}\ and\ \bibinfo {author} {\bibfnamefont {E.}~\bibnamefont
  {Ott}},\ }\bibfield  {title} {\bibinfo {title} {Adaptive synchronization of
  dynamics on evolving complex networks},\ }\href@noop {} {\bibfield  {journal}
  {\bibinfo  {journal} {Phys. Rev. Lett.}\ }\textbf {\bibinfo {volume} {100}},\
  \bibinfo {pages} {114101} (\bibinfo {year} {2008})}\BibitemShut {NoStop}%
\bibitem [{\citenamefont {Lu}\ and\ \citenamefont
  {Qin}(2009)}]{lu2009adaptive}%
  \BibitemOpen
  \bibfield  {author} {\bibinfo {author} {\bibfnamefont {X.~B.}\ \bibnamefont
  {Lu}}\ and\ \bibinfo {author} {\bibfnamefont {B.~Z.}\ \bibnamefont {Qin}},\
  }\bibfield  {title} {\bibinfo {title} {Adaptive cluster synchronization in
  complex dynamical networks},\ }\href@noop {} {\bibfield  {journal} {\bibinfo
  {journal} {Phys. Lett. A}\ }\textbf {\bibinfo {volume} {373}},\ \bibinfo
  {pages} {3650} (\bibinfo {year} {2009})}\BibitemShut {NoStop}%
\bibitem [{\citenamefont {Berner}\ \emph {et~al.}(2020)\citenamefont {Berner},
  \citenamefont {Sawicki},\ and\ \citenamefont {Sch{\"o}ll}}]{berner2020birth}%
  \BibitemOpen
  \bibfield  {author} {\bibinfo {author} {\bibfnamefont {R.}~\bibnamefont
  {Berner}}, \bibinfo {author} {\bibfnamefont {J.}~\bibnamefont {Sawicki}},\
  and\ \bibinfo {author} {\bibfnamefont {E.}~\bibnamefont {Sch{\"o}ll}},\
  }\bibfield  {title} {\bibinfo {title} {Birth and stabilization of phase
  clusters by multiplexing of adaptive networks},\ }\href@noop {} {\bibfield
  {journal} {\bibinfo  {journal} {Phys. Rev. Lett.}\ }\textbf {\bibinfo
  {volume} {124}},\ \bibinfo {pages} {088301} (\bibinfo {year}
  {2020})}\BibitemShut {NoStop}%
\bibitem [{\citenamefont {Berner}\ \emph {et~al.}(2021)\citenamefont {Berner},
  \citenamefont {Vock}, \citenamefont {Sch{\"o}ll},\ and\ \citenamefont
  {Yanchuk}}]{berner2021desynchronization}%
  \BibitemOpen
  \bibfield  {author} {\bibinfo {author} {\bibfnamefont {R.}~\bibnamefont
  {Berner}}, \bibinfo {author} {\bibfnamefont {S.}~\bibnamefont {Vock}},
  \bibinfo {author} {\bibfnamefont {E.}~\bibnamefont {Sch{\"o}ll}},\ and\
  \bibinfo {author} {\bibfnamefont {S.}~\bibnamefont {Yanchuk}},\ }\bibfield
  {title} {\bibinfo {title} {Desynchronization transitions in adaptive
  networks},\ }\href@noop {} {\bibfield  {journal} {\bibinfo  {journal} {Phys.
  Rev. Lett.}\ }\textbf {\bibinfo {volume} {126}},\ \bibinfo {pages} {028301}
  (\bibinfo {year} {2021})}\BibitemShut {NoStop}%
\bibitem [{\citenamefont {Lehnert}\ \emph {et~al.}(2014)\citenamefont
  {Lehnert}, \citenamefont {H{\"o}vel}, \citenamefont {Selivanov},
  \citenamefont {Fradkov},\ and\ \citenamefont
  {Sch{\"o}ll}}]{lehnert2014controlling}%
  \BibitemOpen
  \bibfield  {author} {\bibinfo {author} {\bibfnamefont {J.}~\bibnamefont
  {Lehnert}}, \bibinfo {author} {\bibfnamefont {P.}~\bibnamefont {H{\"o}vel}},
  \bibinfo {author} {\bibfnamefont {A.}~\bibnamefont {Selivanov}}, \bibinfo
  {author} {\bibfnamefont {A.}~\bibnamefont {Fradkov}},\ and\ \bibinfo {author}
  {\bibfnamefont {E.}~\bibnamefont {Sch{\"o}ll}},\ }\bibfield  {title}
  {\bibinfo {title} {Controlling cluster synchronization by adapting the
  topology},\ }\href@noop {} {\bibfield  {journal} {\bibinfo  {journal} {Phys.
  Rev. E}\ }\textbf {\bibinfo {volume} {90}},\ \bibinfo {pages} {042914}
  (\bibinfo {year} {2014})}\BibitemShut {NoStop}%
\bibitem [{\citenamefont {Schr{\"o}der}\ \emph {et~al.}(2016)\citenamefont
  {Schr{\"o}der}, \citenamefont {Chakraborty}, \citenamefont {Witthaut},
  \citenamefont {Nagler},\ and\ \citenamefont
  {Timme}}]{schroder2016interaction}%
  \BibitemOpen
  \bibfield  {author} {\bibinfo {author} {\bibfnamefont {M.}~\bibnamefont
  {Schr{\"o}der}}, \bibinfo {author} {\bibfnamefont {S.}~\bibnamefont
  {Chakraborty}}, \bibinfo {author} {\bibfnamefont {D.}~\bibnamefont
  {Witthaut}}, \bibinfo {author} {\bibfnamefont {J.}~\bibnamefont {Nagler}},\
  and\ \bibinfo {author} {\bibfnamefont {M.}~\bibnamefont {Timme}},\ }\bibfield
   {title} {\bibinfo {title} {Interaction control to synchronize
  non-synchronizable networks},\ }\href@noop {} {\bibfield  {journal} {\bibinfo
   {journal} {Sci. Rep.}\ }\textbf {\bibinfo {volume} {6}},\ \bibinfo {pages}
  {37142} (\bibinfo {year} {2016})}\BibitemShut {NoStop}%
\bibitem [{\citenamefont {Liu}\ \emph {et~al.}(2011)\citenamefont {Liu},
  \citenamefont {Slotine},\ and\ \citenamefont
  {Barab{\'a}si}}]{liu2011controllability}%
  \BibitemOpen
  \bibfield  {author} {\bibinfo {author} {\bibfnamefont {Y.-Y.}\ \bibnamefont
  {Liu}}, \bibinfo {author} {\bibfnamefont {J.-J.}\ \bibnamefont {Slotine}},\
  and\ \bibinfo {author} {\bibfnamefont {A.-L.}\ \bibnamefont {Barab{\'a}si}},\
  }\bibfield  {title} {\bibinfo {title} {Controllability of complex networks},\
  }\href@noop {} {\bibfield  {journal} {\bibinfo  {journal} {Nature}\ }\textbf
  {\bibinfo {volume} {473}},\ \bibinfo {pages} {167} (\bibinfo {year}
  {2011})}\BibitemShut {NoStop}%
\bibitem [{\citenamefont {Yan}\ \emph {et~al.}(2012)\citenamefont {Yan},
  \citenamefont {Ren}, \citenamefont {Lai}, \citenamefont {Lai},\ and\
  \citenamefont {Li}}]{yan2012controlling}%
  \BibitemOpen
  \bibfield  {author} {\bibinfo {author} {\bibfnamefont {G.}~\bibnamefont
  {Yan}}, \bibinfo {author} {\bibfnamefont {J.}~\bibnamefont {Ren}}, \bibinfo
  {author} {\bibfnamefont {Y.-C.}\ \bibnamefont {Lai}}, \bibinfo {author}
  {\bibfnamefont {C.-H.}\ \bibnamefont {Lai}},\ and\ \bibinfo {author}
  {\bibfnamefont {B.}~\bibnamefont {Li}},\ }\bibfield  {title} {\bibinfo
  {title} {Controlling complex networks: How much energy is needed?},\
  }\href@noop {} {\bibfield  {journal} {\bibinfo  {journal} {Phys. Rev. Lett.}\
  }\textbf {\bibinfo {volume} {108}},\ \bibinfo {pages} {218703} (\bibinfo
  {year} {2012})}\BibitemShut {NoStop}%
\bibitem [{\citenamefont {Yuan}\ \emph {et~al.}(2013)\citenamefont {Yuan},
  \citenamefont {Zhao}, \citenamefont {Di}, \citenamefont {Wang},\ and\
  \citenamefont {Lai}}]{yuan2013exact}%
  \BibitemOpen
  \bibfield  {author} {\bibinfo {author} {\bibfnamefont {Z.}~\bibnamefont
  {Yuan}}, \bibinfo {author} {\bibfnamefont {C.}~\bibnamefont {Zhao}}, \bibinfo
  {author} {\bibfnamefont {Z.}~\bibnamefont {Di}}, \bibinfo {author}
  {\bibfnamefont {W.-X.}\ \bibnamefont {Wang}},\ and\ \bibinfo {author}
  {\bibfnamefont {Y.-C.}\ \bibnamefont {Lai}},\ }\bibfield  {title} {\bibinfo
  {title} {Exact controllability of complex networks},\ }\href@noop {}
  {\bibfield  {journal} {\bibinfo  {journal} {Nat. Commun.}\ }\textbf {\bibinfo
  {volume} {4}} (\bibinfo {year} {2013})}\BibitemShut {NoStop}%
\bibitem [{\citenamefont {Menichetti}\ \emph {et~al.}(2014)\citenamefont
  {Menichetti}, \citenamefont {Dall'Asta},\ and\ \citenamefont
  {Bianconi}}]{menichetti2014network}%
  \BibitemOpen
  \bibfield  {author} {\bibinfo {author} {\bibfnamefont {G.}~\bibnamefont
  {Menichetti}}, \bibinfo {author} {\bibfnamefont {L.}~\bibnamefont
  {Dall'Asta}},\ and\ \bibinfo {author} {\bibfnamefont {G.}~\bibnamefont
  {Bianconi}},\ }\bibfield  {title} {\bibinfo {title} {Network controllability
  is determined by the density of low in-degree and out-degree nodes},\
  }\href@noop {} {\bibfield  {journal} {\bibinfo  {journal} {Phys. Rev. Lett.}\
  }\textbf {\bibinfo {volume} {113}},\ \bibinfo {pages} {078701} (\bibinfo
  {year} {2014})}\BibitemShut {NoStop}%
\bibitem [{\citenamefont {Gao}\ \emph {et~al.}(2014)\citenamefont {Gao},
  \citenamefont {Liu}, \citenamefont {D'Souza},\ and\ \citenamefont
  {Barab{\'a}si}}]{gao2014target}%
  \BibitemOpen
  \bibfield  {author} {\bibinfo {author} {\bibfnamefont {J.}~\bibnamefont
  {Gao}}, \bibinfo {author} {\bibfnamefont {Y.-Y.}\ \bibnamefont {Liu}},
  \bibinfo {author} {\bibfnamefont {R.~M.}\ \bibnamefont {D'Souza}},\ and\
  \bibinfo {author} {\bibfnamefont {A.-L.}\ \bibnamefont {Barab{\'a}si}},\
  }\bibfield  {title} {\bibinfo {title} {Target control of complex networks},\
  }\href@noop {} {\bibfield  {journal} {\bibinfo  {journal} {Nat. Commun.}\
  }\textbf {\bibinfo {volume} {5}} (\bibinfo {year} {2014})}\BibitemShut
  {NoStop}%
\bibitem [{\citenamefont {Liu}\ and\ \citenamefont
  {Barab{\'a}si}(2016)}]{liu2016control}%
  \BibitemOpen
  \bibfield  {author} {\bibinfo {author} {\bibfnamefont {Y.-Y.}\ \bibnamefont
  {Liu}}\ and\ \bibinfo {author} {\bibfnamefont {A.-L.}\ \bibnamefont
  {Barab{\'a}si}},\ }\bibfield  {title} {\bibinfo {title} {Control principles
  of complex systems},\ }\href@noop {} {\bibfield  {journal} {\bibinfo
  {journal} {Rev. Mod. Phys.}\ }\textbf {\bibinfo {volume} {88}},\ \bibinfo
  {pages} {035006} (\bibinfo {year} {2016})}\BibitemShut {NoStop}%
\bibitem [{\citenamefont {Lynn}\ and\ \citenamefont
  {Bassett}(2019)}]{lynn2019physics}%
  \BibitemOpen
  \bibfield  {author} {\bibinfo {author} {\bibfnamefont {C.~W.}\ \bibnamefont
  {Lynn}}\ and\ \bibinfo {author} {\bibfnamefont {D.~S.}\ \bibnamefont
  {Bassett}},\ }\bibfield  {title} {\bibinfo {title} {The physics of brain
  network structure, function and control},\ }\href@noop {} {\bibfield
  {journal} {\bibinfo  {journal} {Nat. Rev. Phys.}\ }\textbf {\bibinfo {volume}
  {1}},\ \bibinfo {pages} {318} (\bibinfo {year} {2019})}\BibitemShut {NoStop}%
\bibitem [{\citenamefont {Baggio}\ \emph {et~al.}(2021)\citenamefont {Baggio},
  \citenamefont {Bassett},\ and\ \citenamefont {Pasqualetti}}]{baggio2021data}%
  \BibitemOpen
  \bibfield  {author} {\bibinfo {author} {\bibfnamefont {G.}~\bibnamefont
  {Baggio}}, \bibinfo {author} {\bibfnamefont {D.~S.}\ \bibnamefont
  {Bassett}},\ and\ \bibinfo {author} {\bibfnamefont {F.}~\bibnamefont
  {Pasqualetti}},\ }\bibfield  {title} {\bibinfo {title} {Data-driven control
  of complex networks},\ }\href@noop {} {\bibfield  {journal} {\bibinfo
  {journal} {Nat. Commun.}\ }\textbf {\bibinfo {volume} {12}},\ \bibinfo
  {pages} {1429} (\bibinfo {year} {2021})}\BibitemShut {NoStop}%
\bibitem [{\citenamefont {Cowan}\ \emph {et~al.}(2012)\citenamefont {Cowan},
  \citenamefont {Chastain}, \citenamefont {Vilhena}, \citenamefont
  {Freudenberg},\ and\ \citenamefont {Bergstrom}}]{cowan2012nodal}%
  \BibitemOpen
  \bibfield  {author} {\bibinfo {author} {\bibfnamefont {N.~J.}\ \bibnamefont
  {Cowan}}, \bibinfo {author} {\bibfnamefont {E.~J.}\ \bibnamefont {Chastain}},
  \bibinfo {author} {\bibfnamefont {D.~A.}\ \bibnamefont {Vilhena}}, \bibinfo
  {author} {\bibfnamefont {J.~S.}\ \bibnamefont {Freudenberg}},\ and\ \bibinfo
  {author} {\bibfnamefont {C.~T.}\ \bibnamefont {Bergstrom}},\ }\bibfield
  {title} {\bibinfo {title} {Nodal dynamics, not degree distributions,
  determine the structural controllability of complex networks},\ }\href@noop
  {} {\bibfield  {journal} {\bibinfo  {journal} {PloS one}\ }\textbf {\bibinfo
  {volume} {7}},\ \bibinfo {pages} {e38398} (\bibinfo {year}
  {2012})}\BibitemShut {NoStop}%
\bibitem [{\citenamefont {Gates}\ and\ \citenamefont
  {Rocha}(2016)}]{gates2016control}%
  \BibitemOpen
  \bibfield  {author} {\bibinfo {author} {\bibfnamefont {A.~J.}\ \bibnamefont
  {Gates}}\ and\ \bibinfo {author} {\bibfnamefont {L.~M.}\ \bibnamefont
  {Rocha}},\ }\bibfield  {title} {\bibinfo {title} {Control of complex networks
  requires both structure and dynamics},\ }\href@noop {} {\bibfield  {journal}
  {\bibinfo  {journal} {Scientific reports}\ }\textbf {\bibinfo {volume} {6}},\
  \bibinfo {pages} {24456} (\bibinfo {year} {2016})}\BibitemShut {NoStop}%
\bibitem [{\citenamefont {Kiss}\ \emph {et~al.}(2007)\citenamefont {Kiss},
  \citenamefont {Rusin}, \citenamefont {Kori},\ and\ \citenamefont
  {Hudson}}]{kiss2007engineering}%
  \BibitemOpen
  \bibfield  {author} {\bibinfo {author} {\bibfnamefont {I.~Z.}\ \bibnamefont
  {Kiss}}, \bibinfo {author} {\bibfnamefont {C.~G.}\ \bibnamefont {Rusin}},
  \bibinfo {author} {\bibfnamefont {H.}~\bibnamefont {Kori}},\ and\ \bibinfo
  {author} {\bibfnamefont {J.~L.}\ \bibnamefont {Hudson}},\ }\bibfield  {title}
  {\bibinfo {title} {Engineering complex dynamical structures: Sequential
  patterns and desynchronization},\ }\href@noop {} {\bibfield  {journal}
  {\bibinfo  {journal} {Science}\ }\textbf {\bibinfo {volume} {316}},\ \bibinfo
  {pages} {1886} (\bibinfo {year} {2007})}\BibitemShut {NoStop}%
\bibitem [{\citenamefont {Liu}\ and\ \citenamefont
  {Chen}(2011)}]{liu2011cluster}%
  \BibitemOpen
  \bibfield  {author} {\bibinfo {author} {\bibfnamefont {X.}~\bibnamefont
  {Liu}}\ and\ \bibinfo {author} {\bibfnamefont {T.}~\bibnamefont {Chen}},\
  }\bibfield  {title} {\bibinfo {title} {Cluster synchronization in directed
  networks via intermittent pinning control},\ }\href@noop {} {\bibfield
  {journal} {\bibinfo  {journal} {IEEE Trans. Neural Netw.}\ }\textbf {\bibinfo
  {volume} {22}},\ \bibinfo {pages} {1009} (\bibinfo {year}
  {2011})}\BibitemShut {NoStop}%
\bibitem [{\citenamefont {Sun}\ and\ \citenamefont
  {Motter}(2013)}]{sun2013controllability}%
  \BibitemOpen
  \bibfield  {author} {\bibinfo {author} {\bibfnamefont {J.}~\bibnamefont
  {Sun}}\ and\ \bibinfo {author} {\bibfnamefont {A.~E.}\ \bibnamefont
  {Motter}},\ }\bibfield  {title} {\bibinfo {title} {Controllability transition
  and nonlocality in network control},\ }\href@noop {} {\bibfield  {journal}
  {\bibinfo  {journal} {Phys. Rev. Lett.}\ }\textbf {\bibinfo {volume} {110}},\
  \bibinfo {pages} {208701} (\bibinfo {year} {2013})}\BibitemShut {NoStop}%
\bibitem [{\citenamefont {Cornelius}\ \emph {et~al.}(2013)\citenamefont
  {Cornelius}, \citenamefont {Kath},\ and\ \citenamefont
  {Motter}}]{cornelius2013realistic}%
  \BibitemOpen
  \bibfield  {author} {\bibinfo {author} {\bibfnamefont {S.~P.}\ \bibnamefont
  {Cornelius}}, \bibinfo {author} {\bibfnamefont {W.~L.}\ \bibnamefont
  {Kath}},\ and\ \bibinfo {author} {\bibfnamefont {A.~E.}\ \bibnamefont
  {Motter}},\ }\bibfield  {title} {\bibinfo {title} {Realistic control of
  network dynamics},\ }\href@noop {} {\bibfield  {journal} {\bibinfo  {journal}
  {Nat. Commun.}\ }\textbf {\bibinfo {volume} {4}} (\bibinfo {year}
  {2013})}\BibitemShut {NoStop}%
\bibitem [{\citenamefont {Wells}\ \emph {et~al.}(2015)\citenamefont {Wells},
  \citenamefont {Kath},\ and\ \citenamefont {Motter}}]{wells2015control}%
  \BibitemOpen
  \bibfield  {author} {\bibinfo {author} {\bibfnamefont {D.~K.}\ \bibnamefont
  {Wells}}, \bibinfo {author} {\bibfnamefont {W.~L.}\ \bibnamefont {Kath}},\
  and\ \bibinfo {author} {\bibfnamefont {A.~E.}\ \bibnamefont {Motter}},\
  }\bibfield  {title} {\bibinfo {title} {Control of stochastic and induced
  switching in biophysical networks},\ }\href@noop {} {\bibfield  {journal}
  {\bibinfo  {journal} {Phys. Rev. X}\ }\textbf {\bibinfo {volume} {5}},\
  \bibinfo {pages} {031036} (\bibinfo {year} {2015})}\BibitemShut {NoStop}%
\bibitem [{\citenamefont {Whalen}\ \emph {et~al.}(2015)\citenamefont {Whalen},
  \citenamefont {Brennan}, \citenamefont {Sauer},\ and\ \citenamefont
  {Schiff}}]{whalen2015observability}%
  \BibitemOpen
  \bibfield  {author} {\bibinfo {author} {\bibfnamefont {A.~J.}\ \bibnamefont
  {Whalen}}, \bibinfo {author} {\bibfnamefont {S.~N.}\ \bibnamefont {Brennan}},
  \bibinfo {author} {\bibfnamefont {T.~D.}\ \bibnamefont {Sauer}},\ and\
  \bibinfo {author} {\bibfnamefont {S.~J.}\ \bibnamefont {Schiff}},\ }\bibfield
   {title} {\bibinfo {title} {Observability and controllability of nonlinear
  networks: The role of symmetry},\ }\href@noop {} {\bibfield  {journal}
  {\bibinfo  {journal} {Phys. Rev. X}\ }\textbf {\bibinfo {volume} {5}},\
  \bibinfo {pages} {011005} (\bibinfo {year} {2015})}\BibitemShut {NoStop}%
\bibitem [{\citenamefont {Wang}\ \emph {et~al.}(2016)\citenamefont {Wang},
  \citenamefont {Su}, \citenamefont {Huang}, \citenamefont {Wang},
  \citenamefont {Wang}, \citenamefont {Grebogi},\ and\ \citenamefont
  {Lai}}]{wang2016geometrical}%
  \BibitemOpen
  \bibfield  {author} {\bibinfo {author} {\bibfnamefont {L.-Z.}\ \bibnamefont
  {Wang}}, \bibinfo {author} {\bibfnamefont {R.-Q.}\ \bibnamefont {Su}},
  \bibinfo {author} {\bibfnamefont {Z.-G.}\ \bibnamefont {Huang}}, \bibinfo
  {author} {\bibfnamefont {X.}~\bibnamefont {Wang}}, \bibinfo {author}
  {\bibfnamefont {W.-X.}\ \bibnamefont {Wang}}, \bibinfo {author}
  {\bibfnamefont {C.}~\bibnamefont {Grebogi}},\ and\ \bibinfo {author}
  {\bibfnamefont {Y.-C.}\ \bibnamefont {Lai}},\ }\bibfield  {title} {\bibinfo
  {title} {A geometrical approach to control and controllability of nonlinear
  dynamical networks},\ }\href@noop {} {\bibfield  {journal} {\bibinfo
  {journal} {Nat. Commun.}\ }\textbf {\bibinfo {volume} {7}},\ \bibinfo {pages}
  {11323} (\bibinfo {year} {2016})}\BibitemShut {NoStop}%
\bibitem [{\citenamefont {Za{\~n}udo}\ \emph {et~al.}(2017)\citenamefont
  {Za{\~n}udo}, \citenamefont {Yang},\ and\ \citenamefont
  {Albert}}]{zanudo2017structure}%
  \BibitemOpen
  \bibfield  {author} {\bibinfo {author} {\bibfnamefont {J.~G.~T.}\
  \bibnamefont {Za{\~n}udo}}, \bibinfo {author} {\bibfnamefont
  {G.}~\bibnamefont {Yang}},\ and\ \bibinfo {author} {\bibfnamefont
  {R.}~\bibnamefont {Albert}},\ }\bibfield  {title} {\bibinfo {title}
  {Structure-based control of complex networks with nonlinear dynamics},\
  }\href@noop {} {\bibfield  {journal} {\bibinfo  {journal} {Proc. Natl. Acad.
  Sci. U.S.A.}\ }\textbf {\bibinfo {volume} {114}},\ \bibinfo {pages} {7234}
  (\bibinfo {year} {2017})}\BibitemShut {NoStop}%
\bibitem [{\citenamefont {Sun}\ \emph {et~al.}(2017)\citenamefont {Sun},
  \citenamefont {Leng}, \citenamefont {Lai}, \citenamefont {Grebogi},\ and\
  \citenamefont {Lin}}]{sun2017closed}%
  \BibitemOpen
  \bibfield  {author} {\bibinfo {author} {\bibfnamefont {Y.-Z.}\ \bibnamefont
  {Sun}}, \bibinfo {author} {\bibfnamefont {S.-Y.}\ \bibnamefont {Leng}},
  \bibinfo {author} {\bibfnamefont {Y.-C.}\ \bibnamefont {Lai}}, \bibinfo
  {author} {\bibfnamefont {C.}~\bibnamefont {Grebogi}},\ and\ \bibinfo {author}
  {\bibfnamefont {W.}~\bibnamefont {Lin}},\ }\bibfield  {title} {\bibinfo
  {title} {Closed-loop control of complex networks: A trade-off between time
  and energy},\ }\href@noop {} {\bibfield  {journal} {\bibinfo  {journal}
  {Phys. Rev. Lett.}\ }\textbf {\bibinfo {volume} {119}},\ \bibinfo {pages}
  {198301} (\bibinfo {year} {2017})}\BibitemShut {NoStop}%
\bibitem [{\citenamefont {Hart}\ \emph {et~al.}(2019)\citenamefont {Hart},
  \citenamefont {Zhang}, \citenamefont {Roy},\ and\ \citenamefont
  {Motter}}]{hart2019topological}%
  \BibitemOpen
  \bibfield  {author} {\bibinfo {author} {\bibfnamefont {J.~D.}\ \bibnamefont
  {Hart}}, \bibinfo {author} {\bibfnamefont {Y.}~\bibnamefont {Zhang}},
  \bibinfo {author} {\bibfnamefont {R.}~\bibnamefont {Roy}},\ and\ \bibinfo
  {author} {\bibfnamefont {A.~E.}\ \bibnamefont {Motter}},\ }\bibfield  {title}
  {\bibinfo {title} {Topological control of synchronization patterns: Trading
  symmetry for stability},\ }\href@noop {} {\bibfield  {journal} {\bibinfo
  {journal} {Phys. Rev. Lett.}\ }\textbf {\bibinfo {volume} {122}},\ \bibinfo
  {pages} {058301} (\bibinfo {year} {2019})}\BibitemShut {NoStop}%
\bibitem [{\citenamefont {Morrison}\ and\ \citenamefont
  {Kutz}(2020)}]{morrison2020nonlinear}%
  \BibitemOpen
  \bibfield  {author} {\bibinfo {author} {\bibfnamefont {M.}~\bibnamefont
  {Morrison}}\ and\ \bibinfo {author} {\bibfnamefont {J.~N.}\ \bibnamefont
  {Kutz}},\ }\bibfield  {title} {\bibinfo {title} {Nonlinear control of
  networked dynamical systems},\ }\href@noop {} {\bibfield  {journal} {\bibinfo
   {journal} {IEEE Trans. Netw. Sci.}\ }\textbf {\bibinfo {volume} {8}},\
  \bibinfo {pages} {174} (\bibinfo {year} {2020})}\BibitemShut {NoStop}%
\bibitem [{\citenamefont {Menara}\ \emph {et~al.}(2022)\citenamefont {Menara},
  \citenamefont {Baggio}, \citenamefont {Bassett},\ and\ \citenamefont
  {Pasqualetti}}]{menara2022functional}%
  \BibitemOpen
  \bibfield  {author} {\bibinfo {author} {\bibfnamefont {T.}~\bibnamefont
  {Menara}}, \bibinfo {author} {\bibfnamefont {G.}~\bibnamefont {Baggio}},
  \bibinfo {author} {\bibfnamefont {D.}~\bibnamefont {Bassett}},\ and\ \bibinfo
  {author} {\bibfnamefont {F.}~\bibnamefont {Pasqualetti}},\ }\bibfield
  {title} {\bibinfo {title} {Functional control of oscillator networks},\
  }\href@noop {} {\bibfield  {journal} {\bibinfo  {journal} {Nat. Commun.}\
  }\textbf {\bibinfo {volume} {13}},\ \bibinfo {pages} {4721} (\bibinfo {year}
  {2022})}\BibitemShut {NoStop}%
\bibitem [{\citenamefont {Sanhedrai}\ \emph {et~al.}(2022)\citenamefont
  {Sanhedrai}, \citenamefont {Gao}, \citenamefont {Bashan}, \citenamefont
  {Schwartz}, \citenamefont {Havlin},\ and\ \citenamefont
  {Barzel}}]{sanhedrai2022reviving}%
  \BibitemOpen
  \bibfield  {author} {\bibinfo {author} {\bibfnamefont {H.}~\bibnamefont
  {Sanhedrai}}, \bibinfo {author} {\bibfnamefont {J.}~\bibnamefont {Gao}},
  \bibinfo {author} {\bibfnamefont {A.}~\bibnamefont {Bashan}}, \bibinfo
  {author} {\bibfnamefont {M.}~\bibnamefont {Schwartz}}, \bibinfo {author}
  {\bibfnamefont {S.}~\bibnamefont {Havlin}},\ and\ \bibinfo {author}
  {\bibfnamefont {B.}~\bibnamefont {Barzel}},\ }\bibfield  {title} {\bibinfo
  {title} {Reviving a failed network through microscopic interventions},\
  }\href@noop {} {\bibfield  {journal} {\bibinfo  {journal} {Nat. Phys.}\
  }\textbf {\bibinfo {volume} {18}},\ \bibinfo {pages} {338} (\bibinfo {year}
  {2022})}\BibitemShut {NoStop}%
\bibitem [{\citenamefont {D'Souza}\ \emph {et~al.}(2023)\citenamefont
  {D'Souza}, \citenamefont {Di~Bernardo},\ and\ \citenamefont
  {Liu}}]{d2023controlling}%
  \BibitemOpen
  \bibfield  {author} {\bibinfo {author} {\bibfnamefont {R.~M.}\ \bibnamefont
  {D'Souza}}, \bibinfo {author} {\bibfnamefont {M.}~\bibnamefont
  {Di~Bernardo}},\ and\ \bibinfo {author} {\bibfnamefont {Y.-Y.}\ \bibnamefont
  {Liu}},\ }\bibfield  {title} {\bibinfo {title} {Controlling complex networks
  with complex nodes},\ }\href@noop {} {\bibfield  {journal} {\bibinfo
  {journal} {Nat. Rev. Phys.}\ }\textbf {\bibinfo {volume} {5}},\ \bibinfo
  {pages} {250} (\bibinfo {year} {2023})}\BibitemShut {NoStop}%
\bibitem [{\citenamefont {Roy}\ and\ \citenamefont
  {Thornburg~Jr}(1994)}]{roy1994experimental}%
  \BibitemOpen
  \bibfield  {author} {\bibinfo {author} {\bibfnamefont {R.}~\bibnamefont
  {Roy}}\ and\ \bibinfo {author} {\bibfnamefont {K.~S.}\ \bibnamefont
  {Thornburg~Jr}},\ }\bibfield  {title} {\bibinfo {title} {Experimental
  synchronization of chaotic lasers},\ }\href@noop {} {\bibfield  {journal}
  {\bibinfo  {journal} {Phys. Rev. Lett.}\ }\textbf {\bibinfo {volume} {72}},\
  \bibinfo {pages} {2009} (\bibinfo {year} {1994})}\BibitemShut {NoStop}%
\bibitem [{\citenamefont {Motter}\ \emph {et~al.}(2013)\citenamefont {Motter},
  \citenamefont {Myers}, \citenamefont {Anghel},\ and\ \citenamefont
  {Nishikawa}}]{motter2013spontaneous}%
  \BibitemOpen
  \bibfield  {author} {\bibinfo {author} {\bibfnamefont {A.~E.}\ \bibnamefont
  {Motter}}, \bibinfo {author} {\bibfnamefont {S.~A.}\ \bibnamefont {Myers}},
  \bibinfo {author} {\bibfnamefont {M.}~\bibnamefont {Anghel}},\ and\ \bibinfo
  {author} {\bibfnamefont {T.}~\bibnamefont {Nishikawa}},\ }\bibfield  {title}
  {\bibinfo {title} {Spontaneous synchrony in power-grid networks},\
  }\href@noop {} {\bibfield  {journal} {\bibinfo  {journal} {Nat. Phys.}\
  }\textbf {\bibinfo {volume} {9}},\ \bibinfo {pages} {191} (\bibinfo {year}
  {2013})}\BibitemShut {NoStop}%
\bibitem [{\citenamefont {Zhang}\ \emph {et~al.}(2020)\citenamefont {Zhang},
  \citenamefont {Cao}, \citenamefont {Ouyang},\ and\ \citenamefont
  {Tu}}]{zhangenergy}%
  \BibitemOpen
  \bibfield  {author} {\bibinfo {author} {\bibfnamefont {D.}~\bibnamefont
  {Zhang}}, \bibinfo {author} {\bibfnamefont {Y.}~\bibnamefont {Cao}}, \bibinfo
  {author} {\bibfnamefont {Q.}~\bibnamefont {Ouyang}},\ and\ \bibinfo {author}
  {\bibfnamefont {Y.}~\bibnamefont {Tu}},\ }\bibfield  {title} {\bibinfo
  {title} {The energy cost and optimal design for synchronization of coupled
  molecular oscillators},\ }\href@noop {} {\bibfield  {journal} {\bibinfo
  {journal} {Nat. Phys.}\ }\textbf {\bibinfo {volume} {16}},\ \bibinfo {pages}
  {95} (\bibinfo {year} {2020})}\BibitemShut {NoStop}%
\bibitem [{\citenamefont {Huang}\ \emph {et~al.}(2023)\citenamefont {Huang},
  \citenamefont {Zhang},\ and\ \citenamefont {Braun}}]{huang2023minimal}%
  \BibitemOpen
  \bibfield  {author} {\bibinfo {author} {\bibfnamefont {Y.}~\bibnamefont
  {Huang}}, \bibinfo {author} {\bibfnamefont {Y.}~\bibnamefont {Zhang}},\ and\
  \bibinfo {author} {\bibfnamefont {R.}~\bibnamefont {Braun}},\ }\bibfield
  {title} {\bibinfo {title} {A minimal model of peripheral clocks reveals
  differential circadian re-entrainment in aging},\ }\href@noop {} {\bibfield
  {journal} {\bibinfo  {journal} {Chaos}\ }\textbf {\bibinfo {volume} {33}}
  (\bibinfo {year} {2023})}\BibitemShut {NoStop}%
\bibitem [{kur()}]{kuramotoorderparameter}%
  \BibitemOpen
  \href@noop {} {}\bibinfo {note} {Our energy functional is simply the square
  of the standard Kuramoto order parameter, which makes it differentiable
  everywhere.}\BibitemShut {Stop}%
\bibitem [{SM()}]{SM}%
  \BibitemOpen
  \href@noop {} {}\bibinfo {note} {See Supplemental Material for other choices
  of energy functions, advantage over random switching schemes, and
  tempological control applied to Stuart-Landau oscillators.}\BibitemShut
  {Stop}%
\bibitem [{gra()}]{gradVfootnote}%
  \BibitemOpen
  \href@noop {} {}\bibinfo {note} {In the Kuramoto system, for example, $\nabla
  V = \bm{0}$ at all phase-locked fixed points defined by $\theta_i \in
  \lbrace0, \pi\rbrace$, as well as some \emph{non}-fixed points such as the
  maximally-asynchronous ($V = 1$) ``splayed'' states defined by $\theta_i =
  2\pi(i-1)/n$.}\BibitemShut {Stop}%
\bibitem [{\citenamefont {Billingsley}(2017)}]{billingsley2017probability}%
  \BibitemOpen
  \bibfield  {author} {\bibinfo {author} {\bibfnamefont {P.}~\bibnamefont
  {Billingsley}},\ }\href@noop {} {\emph {\bibinfo {title} {Probability and
  measure}}}\ (\bibinfo  {publisher} {John Wiley \& Sons},\ \bibinfo {year}
  {2017})\BibitemShut {NoStop}%
\bibitem [{\citenamefont {Fisher}\ and\ \citenamefont
  {Tippett}(1928)}]{fisher1928limiting}%
  \BibitemOpen
  \bibfield  {author} {\bibinfo {author} {\bibfnamefont {R.~A.}\ \bibnamefont
  {Fisher}}\ and\ \bibinfo {author} {\bibfnamefont {L.~H.~C.}\ \bibnamefont
  {Tippett}},\ }\bibfield  {title} {\bibinfo {title} {Limiting forms of the
  frequency distribution of the largest or smallest member of a sample},\ }in\
  \href@noop {} {\emph {\bibinfo {booktitle} {Mathematical proceedings of the
  Cambridge philosophical society}}},\ Vol.~\bibinfo {volume} {24}\ (\bibinfo
  {organization} {Cambridge University Press},\ \bibinfo {year} {1928})\ pp.\
  \bibinfo {pages} {180--190}\BibitemShut {NoStop}%
\bibitem [{\citenamefont {David}\ and\ \citenamefont
  {Nagaraja}(2004)}]{david2004order}%
  \BibitemOpen
  \bibfield  {author} {\bibinfo {author} {\bibfnamefont {H.~A.}\ \bibnamefont
  {David}}\ and\ \bibinfo {author} {\bibfnamefont {H.~N.}\ \bibnamefont
  {Nagaraja}},\ }\href@noop {} {\emph {\bibinfo {title} {Order statistics}}}\
  (\bibinfo  {publisher} {John Wiley \& Sons},\ \bibinfo {year}
  {2004})\BibitemShut {NoStop}%
\bibitem [{\citenamefont {Lin}\ and\ \citenamefont
  {Antsaklis}(2009)}]{lin2009stability}%
  \BibitemOpen
  \bibfield  {author} {\bibinfo {author} {\bibfnamefont {H.}~\bibnamefont
  {Lin}}\ and\ \bibinfo {author} {\bibfnamefont {P.~J.}\ \bibnamefont
  {Antsaklis}},\ }\bibfield  {title} {\bibinfo {title} {Stability and
  stabilizability of switched linear systems: a survey of recent results},\
  }\href@noop {} {\bibfield  {journal} {\bibinfo  {journal} {IEEE Transactions
  on Automatic control}\ }\textbf {\bibinfo {volume} {54}},\ \bibinfo {pages}
  {308} (\bibinfo {year} {2009})}\BibitemShut {NoStop}%
\bibitem [{\citenamefont {Axelsson}\ \emph {et~al.}(2008)\citenamefont
  {Axelsson}, \citenamefont {Boccadoro}, \citenamefont {Egerstedt},
  \citenamefont {Valigi},\ and\ \citenamefont {Wardi}}]{axelsson2008optimal}%
  \BibitemOpen
  \bibfield  {author} {\bibinfo {author} {\bibfnamefont {H.}~\bibnamefont
  {Axelsson}}, \bibinfo {author} {\bibfnamefont {M.}~\bibnamefont {Boccadoro}},
  \bibinfo {author} {\bibfnamefont {M.}~\bibnamefont {Egerstedt}}, \bibinfo
  {author} {\bibfnamefont {P.}~\bibnamefont {Valigi}},\ and\ \bibinfo {author}
  {\bibfnamefont {Y.}~\bibnamefont {Wardi}},\ }\bibfield  {title} {\bibinfo
  {title} {Optimal mode-switching for hybrid systems with varying initial
  states},\ }\href@noop {} {\bibfield  {journal} {\bibinfo  {journal}
  {Nonlinear Analysis: Hybrid Systems}\ }\textbf {\bibinfo {volume} {2}},\
  \bibinfo {pages} {765} (\bibinfo {year} {2008})}\BibitemShut {NoStop}%
\end{thebibliography}%

\end{document}